\documentclass[usenatbib]{mn2e}

\usepackage[dvips]{graphicx}
\usepackage{amssymb}
\usepackage{amsmath}
\usepackage{amstext}
\usepackage{multirow} 
\usepackage{lscape}
\usepackage{longtable}

\newcommand{\adfo}{${\rm ADF(O^{2+})}$}

\newcommand{\te}{$T_{e}$}
\newcommand{\nel}{$n_{e}$}

\newcommand{\tf}{$t^{2}$}
\newcommand\ion[2]{#1~{\sc {#2}}\relax}        
\newcommand\ioni[2]{${\rm #1^{#2}}$}           

\newcommand{\cmc}{{\rm cm$^{-3}$}}
\newcommand{\kms}{{\rm km~s$^{-1}$}}
\newcommand{\chb}{{$c({\rm H }\beta)$}}

\newcommand{\ngc}{NGC~7635}
\newcommand{\hii}{H~{\sc ii}}
\newcommand{\hi}{H~{\sc i}}
\newcommand{\ld}{$\lambda$}

\title[Spatial variations in \hii\ regions]{Small-spatial scale variations of nebular properties 
                                               and the abundance discrepancy in three Galactic 
					       \hii\ regions%
					       \thanks{Based on observations made with the 4.2m 
					       William Herschel Telescope (WHT) operated on the 
					       island of La Palma by the Isaac Newton Group in 
					       the Spanish Observatorio del Roque de los 
					       Muchachos of the Instituto de Astrof{\'\i}sica de 
					       Canarias.}}
\author[A. Mesa-Delgado \& C. Esteban]
       {A. Mesa-Delgado\thanks{E-mail: amd@iac.es} and C. Esteban\\
	Instituto de Astrof\'\i sica de Canarias, E-38200 La Laguna, Tenerife, Spain\\ 
	Departamento de Astrof\'\i sica, Universidad de La Laguna, E-38205 La Laguna, 
	Tenerife, Spain\\
       }

\begin{document}

\date{Accepted 2010 March 8.  Received 2010 March 1; in original form 2009 December 14.}
\pagerange{\pageref{firstpage}--\pageref{lastpage}} \pubyear{2009}

\maketitle
\label{firstpage}

\begin{abstract}
 We present results of long-slit spectroscopy in several slit positions that cover different 
 morphological structures of the central parts of three bright Galactic \hii\ regions: M8, M17 
 and \ngc. We study the spatial distributions of a large number of nebular parameters such as the 
 extinction coefficient, line fluxes, physical conditions and ionic abundances at the maximum 
 spatial resolution attainable with our instrumentation. Particularly, our goal is to study the 
 behaviour of the abundance discrepancy factor of \ioni{O}{2+}, \adfo, defined as the 
 logarithmic difference of the \ioni{O}{2+} abundances derived from collisionally excited and 
 recombination lines. We find that the \adfo\ remains fairly constant along the slit positions of 
 M8 and M17. In the case of \ngc, we only detect the \ion{O}{ii} recombination lines in the 
 integrated spectrum along the whole slit, where the \adfo\ reaches a remarkably high value of 
 about 0.59 dex. We compare our results with previous ones obtained for the Orion Nebula. We find 
 several evidences that suggest the presence of a candidate to Herbig-Haro object in M8.      
\end{abstract}


\section{Introduction} \label{intro}
 The study of the elemental abundances in \hii\ regions is an essential tool for our knowledge of 
 the chemical evolution of the universe. Traditionally, ionic abundances relative of the elements 
 heavier than He have been determined from the strong collisionally excited lines (CELs). More 
 than 20 years ago, \citet{french83} obtained the first determination of the 
 \ioni{C}{2+}/\ioni{H}{+} ratio derived from the faint recombination line (RL) \ion{C}{ii} 4267 
 \AA\ for a planetary nebula (PN), finding that it was several orders of magnitude larger than the 
 abundance obtained from the CELs of this ion. Later, this result was confirmed in other PNe 
 \citep[$e.g.$][]{rolastasinska94,mathisliu99}. A similar qualitative result was also found by 
 \cite{peimbertetal93} for the \ioni{O}{2+}/\ioni{H}{+} ratio in the Orion Nebula: the abundances 
 obtained from the flux of the faint RLs were higher than those derived using the standard method 
 based on CELs. Currently, this observational fact is a classical problem in the understanding of 
 the physics of photoionized nebulae known as ``Abundance Discrepancy" (AD) problem. This 
 disagreement is quantified by means of the Abundance Discrepancy Factor (ADF) that can be 
 defined as the ratio, or the logarithmic difference, between abundances of a same ion derived 
 from RLs and CELs. In the case of the \ioni{O}{2+}/\ioni{H}{+} ratio, the ADF has rather similar 
 values between 0.1 and 0.3 dex for extragalactic and Galactic \hii\ regions 
 \citep[$e.g.$][]{garciarojasetal05,garciarojasesteban07,estebanetal09}, while for PNe the ADF 
 shows a much wider range of values, becoming substantially larger in some object 
 \citep[$e.g.$][]{liuetal00,liuetal06,tsamisetal04,tsamisetal08}.

 What causes the AD problem is nowadays debated. On the one hand, the predictions of the 
 temperature fluctuations paradigm proposed by \cite{peimbert67}, and characterized by the mean 
 square of the spatial distribution of temperature --the so-called temperature fluctuations 
 parameter, \tf-- seems to explain the ADF observed in \hii\ regions, as it is argued by 
 \cite{garciarojasesteban07}. Under this scheme, the AD problem is a direct consequence of the 
 different temperature dependence of the emissivities of the lines used: in the case of CELs, it 
 depends exponentially on the electron temperature, \te, of the ionized gas, while the emissivity 
 of RLs has a power law temperature dependence, similar to those of the Balmer lines used as 
 reference to determine the ionic abundance ratio relative to \ioni{H}{+}. On the other hand, the 
 hypothesis suggested by \cite{liuetal00}, where most of the emission of RLs come from a cold 
 hydrogen-poor component immersed in the ambient gas which emits the bulk of CELs, seems to solve 
 the AD problem in PNe with high ADF values. This hypothesis is based on the observed fact that 
 certain PNe contain well resolved H-deficient knots which are strong metallic RL emitters 
 \citep[$e.g.$ Abell 30,][]{harringtonfeibelman84}.  
 
 The existence and origin of the temperature fluctuations is controversial because high values 
 of the \tf\ parameter are not reproduced by standard photoionization models 
 \citep{kingdonferland95,rodriguezgarciarojas10} and additional mechanisms are proposed in order 
 to explain the presence of temperature fluctuations \citep[see revisions of][]%
 {esteban02,peimbertapeimbert06}. In the same way, new scenarios try to find a solution to the AD 
 problem in \hii\ regions putting forward new physical mechanisms. This is the case of the 
 hypothesis presented by \cite{tsamispequignot05} and \cite{stasinskaetal07}. Based on the 
 chemical model for the heavy-elements mixing of \cite{tenoriotagle96}, these authors proposed the 
 presence of two components of different chemical composition and physical conditions in \hii\ 
 regions. The component responsible of most of the emission of RLs consists of cold metal-rich 
 droplets from supernova ejecta still not mixed with the ambient gas of the \hii\ region where 
 most of the CEL emission would be produced. Then, note that assuming a chemically inhomogeneous 
 model in \hii\ regions in order to explain the AD problem, the abundance derived from RLs and 
 CELs would be upper and lower limits, respectively, of the real abundance of the ionized gas 
 \citep{stasinskaetal07}. Recently, a new scenario have been proposed by \cite{ercolano09} based 
 on the existence of high-density quasi-neutral clumps --embedded in the nebular gas ionized by 
 the extreme-ultraviolet (EUV) radiation-- which are ionized mainly by the X-ray emission from 
 the central star. Under this scheme, the CEL emission mainly comes from the region ionized by the 
 EUV radiation (E region), while the RLs are emitted in different proportions from the clumps (X 
 region) and the E region. In this sense, the abundances of the E region would be representative 
 of the nebula and those from CELs would be easier to correct than RL ones. Contrary to the model 
 proposed by \cite{tsamispequignot05} and \cite{stasinskaetal07}, the nebular model of 
 \cite{ercolano09} has homogeneous abundances.
 
 In a previous paper \citep{mesadelgadoetal08} we explored the behaviour of the AD at small 
 spatial scales and its dependence on different nebular parameters and physical conditions in the 
 Orion Nebula in order to shed light on the origin of the AD problem. In that study, we used 
 long-slit spectroscopy at spatial scales of 1\farcs2, finding high \adfo\ values related to the 
 presence of Herbig-Haro (HH) objects and temperature spikes at the position of the 
 protoplanetary discs (proplyds). A subsequent detailed analysis of HH~202 was carried out by 
 \cite{mesadelgadoetal09a} using integral field spectroscopy confirming a high \adfo\ value at 
 the main knot of HH~202 in agreement with the results of \cite{mesadelgadoetal08}. Another 
 important result of \cite{mesadelgadoetal09a} was the obtention --for the first time in an 
 \hii\ region-- of a map of the Balmer temperature and of the temperature fluctuations in the 
 observed field, finding no correlation between the \adfo\ and the \tf\ parameter.
    
 Following the same goals and methodology of \cite{mesadelgadoetal08}, in this paper we have used 
 long-slit spectroscopy at intermediate spectral resolution in order to study the spatial 
 distribution of the \adfo\ and other main nebular properties as well as their relation with the 
 local morphological structures ($e.g.$ density condensations, ionization fronts or HH objects) in 
 other bright Galactic \hii\ regions, namely, M8, M17 and \ngc. 

 After the Orion Nebula, M8 and M17 are probably the most studied Galactic \hii\ regions. M8 forms 
 a blister of photoionized material on the surface of a giant molecular cloud. Near the optical 
 center the region with the highest surface brightness of the nebula is found, the Hourglass (HG) 
 region. This region is mainly ionized by the O star Herschel~36 (Her~36), while the stars 
 HD~165052 and 9~Sgr ionize the rest of the nebula \citep{woodwardetal86}. M17 is a cavity with a 
 V shape and the V opening in the line of sight. The main ionization source of M17 is a group of 
 O3-O4 stars, which belong to the open cluster NGC~6618, located in the dark bay of the nebula 
 \citep{hansonconti95}. A singular characteristic of M17 is its rather high ionization degree 
 --\ioni{O}{2+}/\ioni{O}{+} ratio-- in comparasion with other Galactic \hii\ regions. The chemical 
 composition of M8 and M17 have been widely studied by several authors in all spectral ranges from 
 low to high spectral resolution \citep[$e.g.$][]{rubin69,peimbertcostero69,sanchezpeimbert91,%
 rodriguez99b,estebanetal98,tsamisetal03}. Based on high resolution and deep echelle 
 spectrophotometry, \cite{garciarojasetal07} have provided a complete revision of the chemical 
 abundances of these \hii\ regions using CELs and RLs. Nevertheless, we have not found in the 
 literature detailed studies about the spatial behaviour of the nebular properties of these 
 regions, excluding that of \cite{peimbertetal92} along of 17 areas of M17 at low spectral 
 resolution. Our third region, \ngc, is not a classical \hii\ region. This nebula is an 
 interstellar bubble formed by the interaction of the stellar wind of the O6.5~IIIf star 
 BD$+$60~2522 with the surrounding interstellar medium. Recently works assume that the ram 
 pressure of the stellar wind is balanced by the surrounding gas pressure due to the similarity 
 between the velocities of the the molecular cloud and the bright nebulosity places of the nebula 
 \citep{christopoulouetal95,mooreetal02b}. The study of physical conditions and chemical 
 abundances in \ngc\ have been restricted to some selected zones \citep{talentdufour79,%
 rodriguez99b,mooreetal02b}. \cite{mooreetal02b} have been the first in exploring the spatial 
 distributions of several bright emission lines --[\ion{O}{iii}] 5007 \AA, H$\alpha$ and 
 [\ion{N}{ii}] 6584 \AA-- along the set of knots located at northwest of the central star and the 
 rim of the bubble. These authors also obtained the first density spatial profile along the slit 
 position that covered the knots.    
  \begin{figure*}
   \centering
   \includegraphics[scale=0.7]{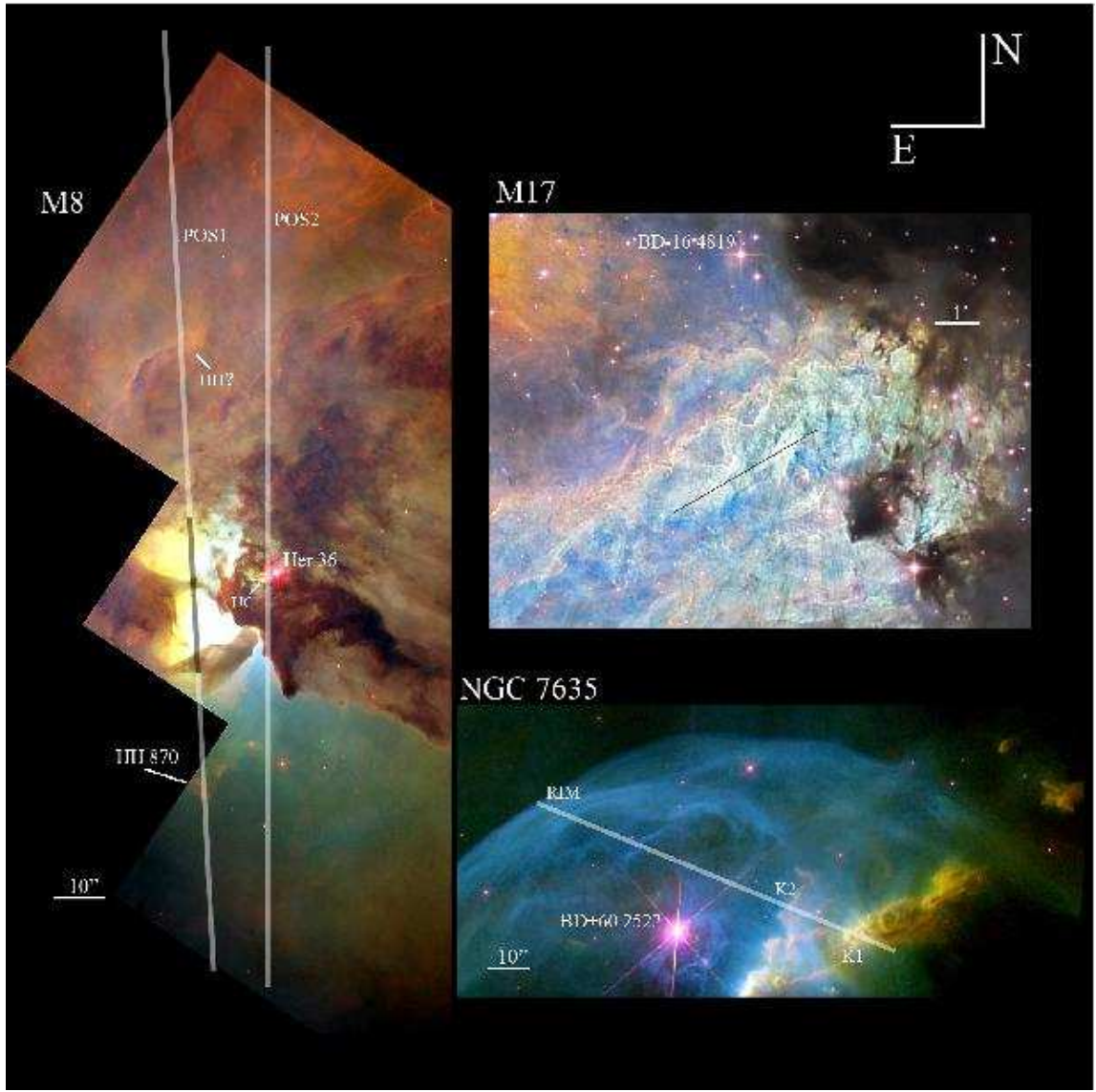} 
   \caption[f1]{Observed slit positions over the central parts of the three Galactic \hii\ regions. 
            The slits only show the total extraction area, while the total slit length is 
	    3\farcm8. The positions of the candidates to proplyd \citep{stecklumetal98} and 
	    HH object (see \S\ref{newhh}) are indicated as UC and HH?, respectively. The images 
	    are combinations of exposures taken with different narrow-band filters. For all 
	    regions, emission from [\ion{O}{iii}] is shown in blue, emission from H$\alpha$ is 
	    shown in green and emission from [\ion{S}{ii}] is shown in red. M8 and \ngc\ images 
	    are sections of combinations of WFPC2 images --\cite{caulet97} and 
	    \cite{mooreetal02b}, respectively. The M17 image is a section of the original mosaic 
	    obtained by the amateur astronomer I. de la Cueva Torregrosa.}
   \label{pos}
  \end{figure*}
 
 In Section~\ref{obsred} we describe the observations of the Galactic \hii\ regions, the reduction 
 procedure and the extraction of the one-dimensional spectra. In Section~\ref{lir} we enumerate 
 the selected emission lines and describe the procedure used to measure the fluxes and the 
 extinction correction applied for each nebula. In Section~\ref{phyab} we describe the method used 
 to determine the physical conditions and the ionic abundances from both kinds of lines, CELs and 
 RLs. In Section~\ref{spatial} we present and discuss the spatial distributions along the slit 
 positions of several nebular quantities for each nebula. In Section~\ref{rlngc} we show the 
 physical conditions and the ionic abundances for the individual extractions of \ngc\ as well as 
 discuss the puzzling abundance pattern found in this object. In Section~\ref{discu} we compare 
 the results on the light of the different lineal spatial resolution attained in this paper with 
 that of the observations of \cite{mesadelgadoetal08} in the case of the Orion Nebula. We also 
 present several arguments for the presence of a new HH object in M8 and discuss the possible 
 causes of the high \adfo\ found in \ngc. Finally, in Section~\ref{conclu} we summarize the 
 main conclusions of the paper.  
\section{Observations, data Reduction and extraction of the one-dimensional spectra} 
\label{obsred}
 Long-slit spectra at intermediate spectral resolution were obtained on 18 July 2007 and 11 May 
 2008 using the Intermediate dispersion Spectrograph and Imaging System (ISIS) at the 4.2m William 
 Herschel Telescope (WHT) in the Observatorio del Roque de los Muchachos (La Palma, Spain). 
 Two different CCDs were used at the blue and red arms of the spectrograph: an EEV12 CCD with a 
 configuration 4096$\times$2048 pixels with a size of 13.5$\mu$m per pixel in the blue arm and a 
 REDPLUS CCD with 4096$\times$2048 pixels with a pixel size of 15$\mu$m in the red arm. The 
 spatial scale was 0\farcs20 pixel$^{-1}$  and 0\farcs22 pixel$^{-1}$ in the blue and red arm, 
 respectively. The slit length was 3\farcm8 and the slit width was fixed to 0\farcs98. The R1200B 
 grating was used in the blue arm and the R316R one in the red arm. These gratings gave an 
 effective spectral resolution of 0.86 and 3.81 \AA\ for the blue and red arms, respectively. The 
 blue spectra covered the spectral range from 4220 to 5080 \AA\ and the red one from 5320 to 8100 
 \AA. The observation nights were photometric and the seeing during both observations was between 
 of 0\farcs5 and 0\farcs8.

 The Galactic \hii\ region M8 and the wind-blown nebula \ngc\ were observed on 18 July 2007. In 
 the case of M8, we observed two slit positions (POS1 and POS2) centered at the HG region with 
 different position angles (see Fig.~\ref{pos}, the HG region is indicated by a grey section over 
 the slit of POS1). These positions were chosen in order to cover different morphological 
 structures such as a candidate to proplyd \citep{stecklumetal98} marked as UC in Fig.~\ref{pos} 
 and the prominent HH~870. Unfortunately, during the analysis of these data, we noticed that 
 probably the proplyd was not covered by the slit as we expected due to the difficulty of 
 possitioning the slit so close to the bright star Her~36. In the case of \ngc, a single slit 
 position was used to cover the bright knots K1 and K2 as well as the rim of the bubble 
 (see Fig.~\ref{pos}). On 11 May 2008 a single slit position was observed over the Galactic 
 \hii\ region M17 covering a high surface brightness zone. For all objects, large and short 
 exposures  were taken at each slit position and spectral range in order to achieve a good 
 signal-to-noise ratio in the faint \ion{C}{ii} and \ion{O}{ii} RLs and to avoid saturation of 
 the brightest emission lines. The journal of observations can be found in Table~\ref{jobs} where 
 we present the coordinates (RA, DEC) and the position angle (PA) of each slit position observed 
 as well as the total exposure times.
  \begin{table*}
   \centering
   \begin{minipage}{110mm}
     \caption{Journal of observations.}
     \label{jobs}
    \begin{tabular}{cccccc}
     \hline
              &        &         &           & \multicolumn{2}{c}{Exposure time (s)}\\     
       Target & RA$^a$ & DEC$^a$ & PA ($^\circ$) &  Blue arm & Red arm \\
     \hline
       M8 POS1 & 18$^h$03$^m$41\fs40 & $-$24$^\circ$22$'$42\farcs7 &    3 &  7$\times$1200,60 & 36$\times$180,60\\
M8 POS2 & 18$^h$03$^m$40\fs42 & $-$24$^\circ$22$'$42\farcs7 &    0 &  7$\times$1200,60 & 36$\times$180,60\\
M17     & 18$^h$20$^m$42\fs98 & $-$16$^\circ$10$'$02\farcs4 &  120 &  5$\times$1800,60 & 15$\times$500,60\\
\ngc\   & 23$^h$20$^m$44\fs36 & $+$61$^\circ$11$'$56\farcs0 & 67.5 &  5$\times$1200,60 &  8$\times$600,60\\

     \hline
    \end{tabular}
    \begin{description}
      \item[$^a$] Coordinates of the slit center (J2000.0).  
    \end{description}
   \end{minipage}
  \end{table*}
 
 All CCD frames were reduced using the standard {\sc iraf}\footnote{IRAF is distributed by 
 National Optical Astronomical Observatories, operated by the Associated Universities for 
 Research in Astronomy, under cooperative agreement with the National Science Foundation} 
 {\sc twodspec} reduction package to perform bias correction, flat-fielding, cosmic-ray rejection, 
 wavelength and flux calibration. The wavelength calibration was carried out with a CuNe+CuAr 
 lamp. The absolute flux calibration was achieved by observations of the standard stars 
 BD$+$33~2642, BD$+$28~4211 and BD$+$25~4655 for the first night and  BD$+$25~3941, Feige~34 and 
 BD$+$33~2642 for the second one. The error of the flux calibration is of the order of 5 per cent.
 
 The extraction of the one-dimensional spectra was done using an {\sc iraf} script based on 
 {\sc apall} task, following the procedure explained in \cite{mesadelgadoetal08}. We extracted 
 the apertures for each region applying the bidimensional fit to the spectra of a standard star 
 (with positional coordinates more similars to the object) used for flux calibration on the 
 bidimensional image. In all cases, we adjusted a third-order Chebyshev polynomial obtaining a 
 typical $rms$ between 0.05 and 0.08 pixels. The slit center in the red arm was some pixels 
 displaced with respect to the slit center in the blue arm; this effect was also corrected in the 
 extraction procedure, and later verified from the alignment of H$\alpha$ and H$\beta$ spatial 
 profiles, ensuring the same spatial coverage in both ranges. We also discarded apertures located 
 at the edges of the CCD.  
 
 For M8 we extracted apertures of 6 pixels in the blue arm, and 5.45 pixels in the red one, in 
 the spatial direction, which corresponds to an effective spatial resolution of 1\farcs2. This 
 size was chosen as a compromise to have the maximum attainable spatial resolution (only slightly 
 larger than the mean seeing of the night) and a good signal-to-noise ratio in the auroral lines 
 of [\ion{O}{iii}] and [\ion{N}{ii}]. However, larger extractions of 4\farcs8 angular size were 
 necessary to achieve a good flux measurement in the fainter \ion{O}{ii} RLs. Then, for the slit 
 position 1 and considering the discarded apertures, we have obtained a final number of 150 
 apertures with a spatial resolution of 1\farcs2$\times$0\farcs98 and 38 apertures with a 
 resolution of 4\farcs8$\times$0\farcs98 for the measurements of \ion{O}{ii} RLs. Similarly, a 
 total number of 146 and 37 apertures were extracted, respectively, for the slit position 2. The 
 smaller number of apertures obtained in this slit position is due to the fact that we discarded 
 4 apertures which were severely contaminated by stellar emission from Her~36.
 
 In the case of \ngc\ slit position, we extracted apertures with an angular size of 3\arcsec. A 
 total of 30 apertures with an area of 3\arcsec$\times$0\farcs98 were used. We also extracted 
 individual integrated spectra of knots K1 and K2, and the rim of the bubble with areas of 
 12\farcs$4\times$0\farcs98, 8\farcs$5\times$0\farcs98 and 10\farcs4$\times$0\farcs98, 
 respectively. We notice the detection of the \ion{C}{ii} 4267 \AA\ RL in the three extractions.
 
 For M17, apertures extracted with an angular size of 1\farcs2 was sufficient to obtain a high 
 signal-to-noise ratio in the auroral lines and RLs. We have obtained a total number of 185 
 apertures extracted with an individual area of 1\farcs2$\times$0\farcs98.
 
 Additionally, for each slit position, we extracted a one-dimensional spectra collapsing the whole 
 slit --the sum of all the individual apertures. These are designated as ``whole slit" spectra. 
 The area covered by the ``whole slit", and therefore the total extraction area, is shown in 
 Fig.~\ref{pos}. 
\section{Emission Line measurements and reddening correction} \label{lir}
 The emission lines considered in our analysis were selected according to the following criteria: 
 \begin{itemize}
  \item \hi\ lines -- H$\alpha$, H$\beta$ and H$\gamma$--, which are used to compute the 
  reddening correction and to re-scale the line flux ratios of the red spectral range with 
  respect to the blue one.
  \item CELs of several species in order to compute the physical conditions such as the auroral 
  lines [\ion{O}{iii}] 4363 \AA\ and [\ion{N}{ii}] 5755 \AA\ used to derive the electron 
  temperatures, [\ion{S}{ii}] 6717, 6731 \AA\ and [\ion{Cl}{iii}] 5718, 5738 \AA\ used to 
  calculate the electron density.
  \item Other CELs needed to derive different ionic abundances (\ioni{N}{+}, \ioni{O}{+}, 
  \ioni{O}{2+}, \ioni{S}{+}, \ioni{S}{2+}, \ioni{Cl}{2+} and \ioni{Ar}{2+}    )   
  \item Faint RLs of \ion{C}{ii} and \ion{O}{ii}, which are used to derive the \ioni{C}{2+} and 
  \ioni{O}{2+} abundances and to compute the abundance discrepancy factor, ADF, for \ioni{O}{2+} 
  (via a comparison with the \ioni{O}{2+} abundances derived from CELs).
 \end{itemize}  
 
 Line fluxes were measured applying a single or a multiple Gaussian profile fit procedure over a 
 local continuum. All these measurements were made with the {\sc splot} routine of the {\sc iraf} 
 package and using our own scripts to automatize the process. Due to the local variations of the 
 continuum around \ion{O}{ii} RLs and their faintness, these lines were measured manually.
  
 Following \cite{mesadelgadoetal08}, to accurately compute the line fluxes we need to define the 
 adjacent continuum of each line using the {\sc splot} routine. For each selected line, we define 
 two small spectral zones at each side of the line as close as possible and free of any spectral 
 feature. Then, the routine fits the continuum between both zones and the line profile to obtain 
 the flux. The observational errors associated with the flux measurements were estimated following 
 the criteria of \cite{mesadelgadoetal08}. The final error of a line flux was computed as 
 the quadratic sum of the error in its flux measurement and the error in the flux calibration. 
 All line fluxes for a given aperture were normalized to an \ion{H}{i} line of reference, 
 H$\beta$ and H$\alpha$ for the blue and red range, respectively.

 The observed fluxes with respect to their \ion{H}{i} line of reference were dereddened using the 
 usual relation,
 \begin{equation}
   \frac{I(\lambda)}{I(\lambda_{ref})} = \frac{F(\lambda)}{F(\lambda_{ref})}
                                        10^{c({\rm H}\beta)\big[f(\lambda)-f(\lambda_{ref})\big]}, 
 \end{equation} 
 where the reddening coefficient, \chb, represents the amount of interstellar extinction, 
 $f(\lambda)$ the adopted extinction curve normalized to $f({\rm H}\beta)=0$, and $\lambda_{ref}$ 
 the \ion{H}{i} line of reference. The reddening coefficient was determined from the comparison 
 of the observed flux ratio of H$\gamma$ and H$\alpha$ with respect to H$\beta$ and the case B 
 theoretical ones computed by \cite{storeyhummer95} for the physical conditions of \te\ $=$ 
 10000 K and \nel\ $=$ 1000 \cmc. The final \chb\ was the weighted average of the values obtained 
 from each line. In appendix~\ref{ap1}, we present the dereddened emission line ratios and their 
 associatted errors of the main emission lines per slit position as well as the observed 
 H$\beta$ flux and the mean \chb\ coefficient (just in the online version).
 
 In all regions we have assumed the extinction law derived by \cite{cardellietal89}, which is 
 parametrized by the ratio of total to selective extinction, $R_V = A_V/E(B-V)$. In the case of 
 M17, \ngc\ and the position 2 of M8, we have used the typical value in the diffuse interstellar 
 medium, $R_V=3.1$. For the position 1 of M8 we have adopted different values of $R_V$ depending 
 on the zones covered by the slit. It is well known that the main ionization source of HG is 
 Her~36 \citep{woodwardetal86} and that it shows a considerably higher extinction than the other 
 zones of M8, which are mainly ionized by 9~Sgr and most of their reddening is due to foreground 
 interstellar dust. Consequently, following \cite{sanchezpeimbert91} we have adopted a $R_V=5.0$ 
 to correct the apertures of position 1 that cover the HG region (the area indicated by a 
 darker grey band in the slit position 1 of M8 shown in Fig.~\ref{pos}) and the typical value 
 $R_V=3.1$ in the other zones. The use of a higher total to selective extinction produces 
 significant changes in the \chb\ values, while the line flux ratios remain almost unaffected by 
 deviations from the classical extinction law. In the case of the ``whole slit" spectra, we have 
 also assumed the typical $R_V$ value for the interstellar medium. Finally, in order to produce a 
 final homogeneous set of dereddened flux ratios, all of them were re-scaled to H$\beta$. The 
 re-scaling factor used in the red spectra was the theoretical H$\alpha$/H$\beta$ ratio for the 
 physical conditions of \te $=$ 10000 K and \nel $=$ 1000 \cmc. The final error associated with 
 the line dereddened fluxes include the uncertainties in the flux measurements, flux calibration 
 and the error propagation in the reddening coefficient.
  \begin{figure*}
   \centering
   \includegraphics[scale=0.6,angle=90]{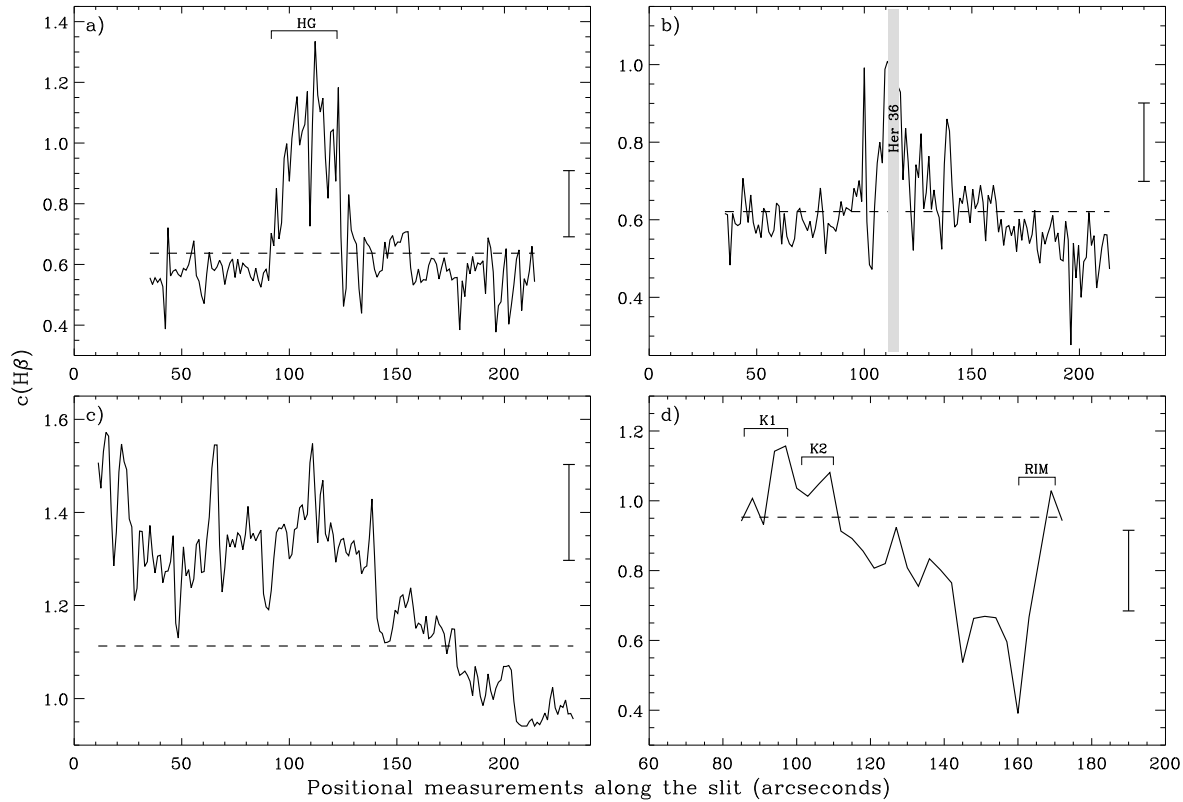} 
   \caption[extinction profiles]{Spatial profiles of the reddening coefficient, \chb, along the 
            slit positions: a) M8 POS1, b) M8 POS2, c) M17 and d) \ngc. The position of the 
	    Hourglass region (HG) as well as of the knots K1, K2 and the rim of the bubble of 
	    \ngc\ are indicated. The grey band in b) corresponds to apertures contaminated by 
	    stellar emission from Her~36. Positional measurements along the slits go from south 
	    to north in the two positions of M8 and from the west to east for M17 and \ngc. The 
	    dashed horizontal line represents the \chb\ value obtained for the ``whole slit" 
	    spectra. The typical error bar is also included.}
   \label{chb}
  \end{figure*}  
 
 In the different panels of Fig.~\ref{chb} we present the spatial profiles of the \chb\ 
 determinations as a function of the positional measurement along the slit in arcseconds. 
 Hereinafter, each positional measurement represents the position on the slit of a given quantity 
 obtained for an individual extraction with origin in the south edge of the original slit in the 
 case of M8, in the northwest edge for M17 and in the southwest edge for \ngc. In all cases, we 
 have found a good agreement between our determinations and those available in the literature, 
 considering that the measurements do not correspond exactly to the same spatial zone and the 
 areas covered by the slits are also different. On the one hand, Figs~\ref{chb}(a) and \ref{chb}(b) 
 show the spatial distribution of \chb\ for slit positions 1 and 2 of M8, respectively, where we 
 can notice higher \chb\ values associated with the HG region and the zones near Her~36. These 
 values are in agreement with previous \chb\ determinations carried out in the HG region by 
 \cite{sanchezpeimbert91}, \cite{estebanetal99b} and \cite{garciarojasetal07}, who found values 
 between 0.85 and 1.0 dex. On the other hand, in Fig.~\ref{chb}(c) we present the spatial 
 distribution of \chb\ along the slit position of M17. We have compared our determinations with 
 those performed by \cite{peimbertetal92}, whose slit positions 3 and 13 approximately coincide 
 with the positional measurements 60\arcsec\ and 125\arcsec, respectively. For those positions, 
 \cite{peimbertetal92} obtained values of about 1.72 and 1.45 dex, which are in agreement with 
 our determinations considering the uncertainties in the \chb\ determination. Fig.~\ref{chb}(d) 
 presents the \chb\ spatial profile of \ngc. In this panel, we can see higher \chb\ values 
 related to the knots and the rim. We have compared our values with previous determinations 
 obtained by \cite{talentdufour79} and \cite{mooreetal02b} at different parts of the nebula. 
 Positions 3 and 5 of \cite{talentdufour79} coincide with the positions of the knot K1 and the 
 rim indicated in Fig.~\ref{chb}(d). \cite{talentdufour79} measured \chb\ values at those 
 positions of about 1.21 and 1.07 dex, respectively, which are in agreement with our 
 determinations. On the other hand, \cite{mooreetal02b} obtained a similar value, 1.25 dex, from 
 their slit position that covers the knots K1 and those located to the south of K1, but measured 
 a value of about 1.46 dex for the rim of the bubble, which is higher than our determination and 
 that of \cite{talentdufour79} at the same position. The typical error in the \chb\ coefficient 
 presented in Fig.~\ref{chb} is of about 0.1 dex for all regions.
 
 The results shown in Fig.~\ref{chb} indicate that an important fraction of the extinction should 
 come from dust located inside the objects, and that the distribution of such absorbing material 
 is not homogeneous.
\section{Physical conditions and chemical abundances} \label{phyab}     
 \subsection{Physical conditions}\label{phy}
  We have determined the physical conditions --electron densities and temperatures-- from the 
  usual CEL ratios and using the {\sc iraf} task {\sc temden} of the {\sc nebular} package 
  \citep{shawdufour95} with updated atomic data \citep[see][]{liuetal00,garciarojasetal05}. We 
  have computed the electron density, \nel, from the [\ion{S}{ii}] 6717/6731 line ratio and the 
  electron temperatures, \te, from the nebular to auroral [\ion{O}{iii}] (4959+5007)/4363 and 
  [\ion{N}{ii}] (6548+6584)/5755 line ratios. Although we detect the [\ion{Cl}{iii}] doublet in 
  several apertures, we do not use these lines to obtain density values due to their large 
  associated errors. The spatial distributions of the physical conditions are presented and 
  discussed for each region in \S\ref{spatial}.  

  Following the same methodology as \cite{mesadelgadoetal08} for the determination of the physical 
  conditions, a representative initial \te\ $=$ 10000 K is assumed in order to derive a first 
  approximation of \nel([\ion{S}{ii}]) --hereinafter \nel. Then, we calculate 
  \te([\ion{O}{iii}]) and \te([\ion{N}{ii}]), and iterate until convergence to compute the finally 
  adopted values using \te([\ion{N}{ii}]) in the density calculations. The errors in the 
  physical conditions were computed by error propagation on the analytical expressions of 
  \nel\ by \cite{castanedaetal92} and those of \te\ given by \cite{osterbrockferland06} 
  (their equations 5.4 and 5.5). Although the expression derived by \cite{castanedaetal92} is only 
  valid to a limited range of densities lower than 10$^4$ \cmc, and uses the old atomic data for 
  \ioni{S}{+} from the compilation by \cite{mendoza83}, it seems adequate for an estimation of 
  the errors in the physical conditions.
 \subsection{Ionic abundances from CELs and RLs} \label{ab}
  We used the {\sc iraf} package {\sc nebular} in order to derive ionic abundances of \ioni{N}{+}, 
  \ioni{O}{+}, \ioni{O}{2+}, \ioni{S}{+}, \ioni{S}{2+}, \ioni{Cl}{2+} and \ioni{Ar}{2+} from CELs.
  We have assumed no temperature fluctuations in the ionized gas (\tf $= 0$) and a two-zone 
  scheme, adopting the \te([\ion{N}{ii}]) to derive the abundances of singly ionized species and 
  \te([\ion{O}{iii}]) in the case of doubly ionized species. The electron density obtained from 
  the[\ion{S}{ii}] line ratio was adopted for all ionic species. The errors in the ionic abundance 
  determinations were calculated as the quadratic sum of the independent contributions of 
  temperature, density, and line flux uncertainties. The spatial distributions of the ionic 
  abundances of some species are presented and discussed for each region in \S\ref{spatial}.
  \begin{figure*}
   \centering
   \includegraphics[scale=0.6,angle=90]{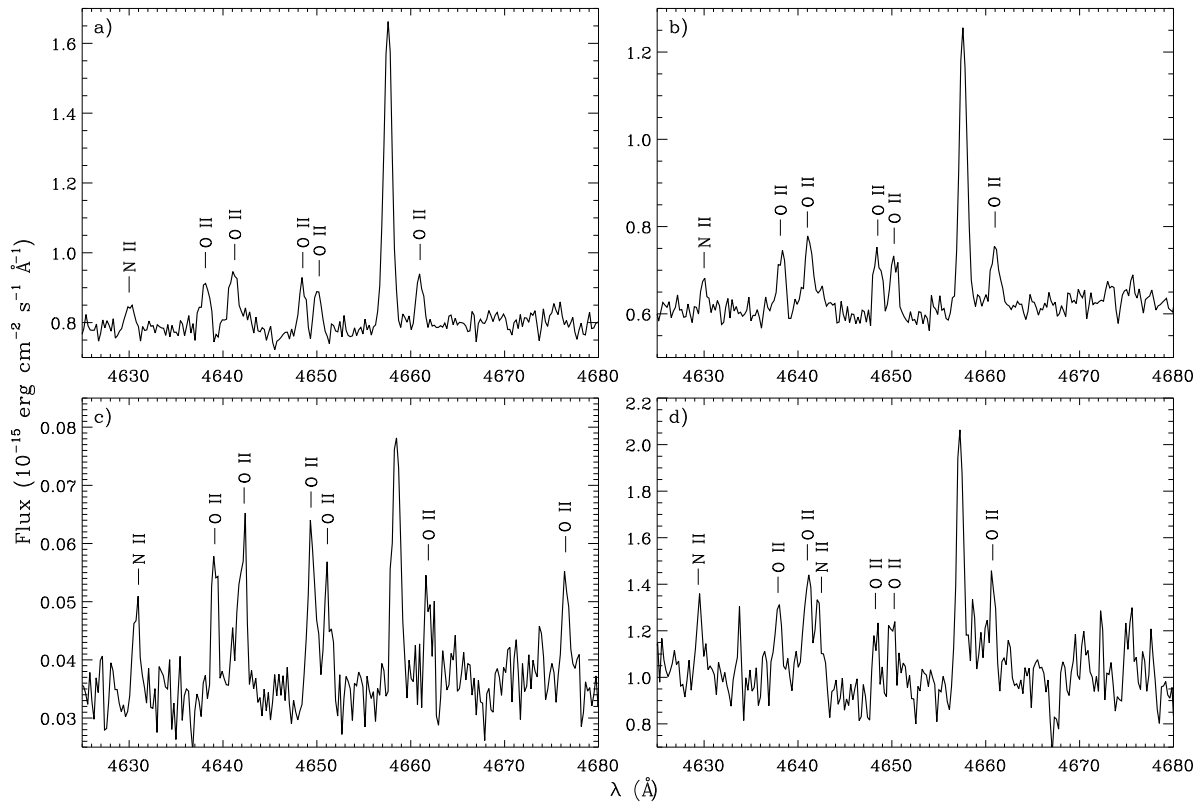} 
   \caption[recl]{Sections of the spectrum around the emission lines of multiplet 1 of 
                  \ion{O}{ii}. Panels a) and b) correspond to apertures of 4\farcs8 wide 
		  extracted from slit positions 1 and 2 of M8, respectively, extending from 
		  positional measurements 137\farcs2 to 142\arcsec. Panel c) corresponds to an 
		  1\farcs2 wide aperture extracted from the slit position of M17, which covers 
		  from 153\farcs4 to 154\farcs6. Panel d) shows a section of the ``whole slit" 
		  spectrum of \ngc.}
   \label{rls}
  \end{figure*}  
  
  The wavelength range covered with the blue arm did not include the bright [\ion{O}{ii}] 3726, 
  3729 \AA\ lines. In the case of \ngc, it was necessary to determine the \ioni{O}{+} abundance 
  (see \S\ref{rlngc}); to do that, we performed a proper subtraction of the telluric emission in 
  the [\ion{O}{ii}] 7320, 7330 \AA\ lines. The subtraction was achieved using re-scaled pure 
  telluric measurements from sky extractions of different sizes of long exposure spectra of a 
  compact \hii\ region observed during the same night \citep[the data of this object were %
  published in][]{martinhernandezetal08}. This procedure adds an uncertainty between 6\% and 8\% 
  higher in the flux measurements of [\ion{O}{ii}] lines. On the other hand, in the cases of M8 
  and M17 we consider not necessary to determine the \ioni{O}{+} abundance for our aims. 

  We have detected and measured pure RLs of the multiplet 1 of \ion{O}{ii} (see Fig.~\ref{rls}) 
  and \ion{C}{ii} 4267 \AA. The abundance of a heavy element X in the ionization state $i+1$, 
  that emits a RL at wavelength \ld\ is given by  
  \begin{equation}
   \frac{N({\rm X}^{i+1})}{N({\rm H}^+)} = %
   \frac{\lambda({\rm \AA)}}{4861} \frac{\alpha_{eff}({\rm H}\beta)}{\alpha_{eff}(\lambda)} %
   \frac{I(\lambda)}{I({\rm H}\beta)},
  \end{equation}
  where $I(\lambda)/I($H$\beta)$  is the dereddened flux ratio, $\alpha_{eff}(\lambda)$ and 
  $\alpha_{eff}($H$\beta)$ are the effective recombination coefficients for the RL and H$\beta$, 
  respectively. Due to the similar temperature dependence of the emissivities of RLs, the 
  $\alpha_{eff}({\rm H}\beta)/\alpha_{eff}(\lambda)$ ratio is almost independent of the adopted 
  temperatures. In our case, under the two-zone scheme, we have assumed \te([\ion{O}{iii}]) to 
  calculate the \ioni{C}{2+} and \ioni{O}{2+} abundances.

  We have detected the \ion{C}{ii} line at 4267 \AA\ in most of the apertures extracted at 
  1\farcs2 in M17 and M8, and in the integrated spectra for the ``whole slit", the knots and the 
  rim of the bubble of \ngc\ (see \S\ref{rlngc}). \ioni{C}{2+} abundances from RLs have been 
  calculated using the atomic data of \cite{daveyetal00}.
  
  The \ioni{O}{2+} abundances were derived when at least four lines of multiplet 1 were measured 
  in a given one-dimensional spectrum using the effective recombination coefficients from 
  \cite{storey94} and the same method that is detailed in \cite{estebanetal98}. We have 
  determined the \ioni{O}{2+} abundances correcting for the departure from local thermodinamyc 
  equilibrium (LTE) of the upper levels of the transitions of multiplet 1 of \ion{O}{ii} for 
  densities lower than 10000 \cmc\ using the empirical formulation proposed by 
  \cite{apeimbertpeimbert05}. Abundances determined assuming LTE for the population of the levels 
  do not differ more than a 2\% from the NLTE ones because we use several of the brightest lines 
  of the multiplet. We have computed the \ioni{O}{2+}/\ioni{H}{+} ratio for 89\% of the apertures 
  extracted at 1\farcs2 in M17, for all the apertures extracted at 4\farcs8 in both slit 
  positions of M8, and in the ``whole slit" spectrum of \ngc\ (see \S\ref{rlngc}).
\section{Spatial profiles along the slit positions} \label{spatial}     
 In the following sections we present the spatial profiles of several nebular properties along 
 the slit positions of each \hii\ region. The selected parameters were: \nel, \te([\ion{N}{ii}]), 
 \te([\ion{O}{iii}]), the flux of several selected lines ([\ion{Fe}{iii}] 4658 \AA, \ion{C}{ii} 
 4267 \AA, \ion{O}{ii} 4649 \AA\ and [\ion{O}{iii}] 4959 \AA), and the \ioni{O}{2+} abundances 
 obtained from CELs and RLs.
 \subsection{M8}\label{m8}
\begin{figure*}
   \centering
   \includegraphics[scale=0.6,angle=90]{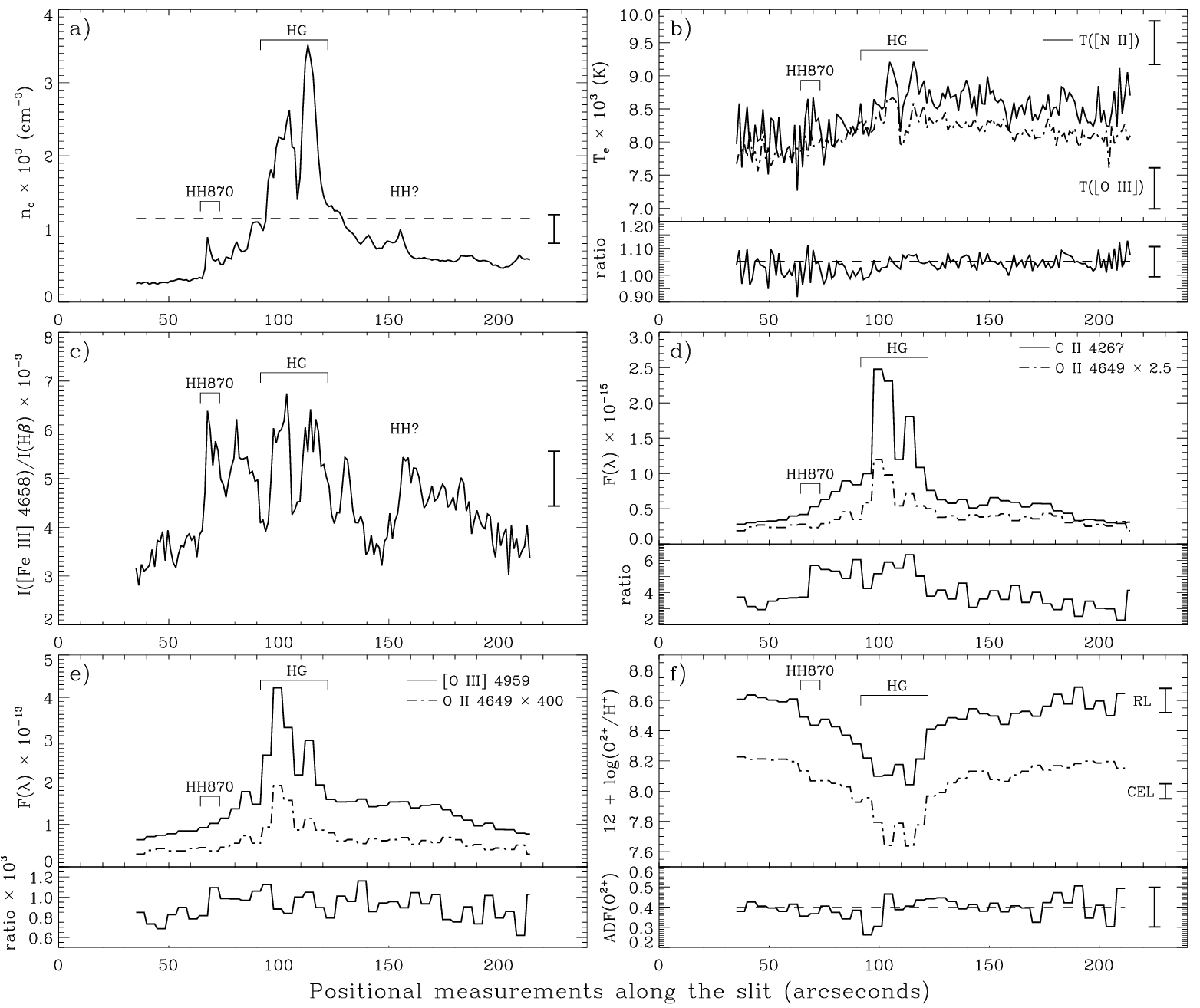} 
   \caption[Spatial profiles of M8 POS1]{Spatial profiles of several nebular parameters along 
            the slit position 1 of M8. Positional measurements along the slit go from south to 
	    north (see Fig.~\ref{pos}). The position of the Hourglass (HG) region and the HH 
	    object 870 are indicated as well as the typical error of some variables. The possible 
	    candidate to HH object (see \S\ref{newhh}) is also marked as HH? in panels (a) and 
	    (c). The horizontal dashed line in some panels indicates the value obtained for the 
	    ``whole slit" spectrum. (a) Profile of \nel; (b) top: profiles of \te([\ion{N}{ii}]) 
	    (solid line) and \te([\ion{O}{iii}]) (dash-dotted line), botton: profile of the 
	    \te([\ion{N}{ii}])/\te([\ion{O}{iii}]) ratio; (c) profile of the dereddened flux ratio
	    of [\ion{Fe}{iii}] 4658 \AA; (d) top: observed flux of \ion{C}{ii} 4267 \AA\ (solid 
	    line) and \ion{O}{ii} 4649 \AA\ (dash-dotted line), botton: profile of the 
	    F(\ion{C}{ii} 4267 \AA)/F(\ion{O}{ii} 4649 \AA) ratio; (e) top: observed flux of 
	    [\ion{O}{iii}] 4959 \AA\ (solid line) and \ion{O}{ii} 4649 \AA\ (dash-dotted line), 
	    botton: profile of the F([\ion{O}{iii}] 4959 \AA)/F(\ion{O}{ii} 4649 \AA) ratio; (f) 
	    top: \ioni{O}{2+} abundances from RLs (solid line) and CELs (dash-dotted line), 
	    botton: profile of \adfo. All observed fluxes have units of erg~cm$^{-2}$~s$^{-1}$. 
	    Note that from panel a) to c) the profiles correspond to extractions of 1\farcs2. 
	    From panels d) to f) the spatial increments are 4\farcs8 wide.}
   \label{m8p1} 
  \end{figure*}

\begin{figure*}
   \centering
   \includegraphics[scale=0.6,angle=90]{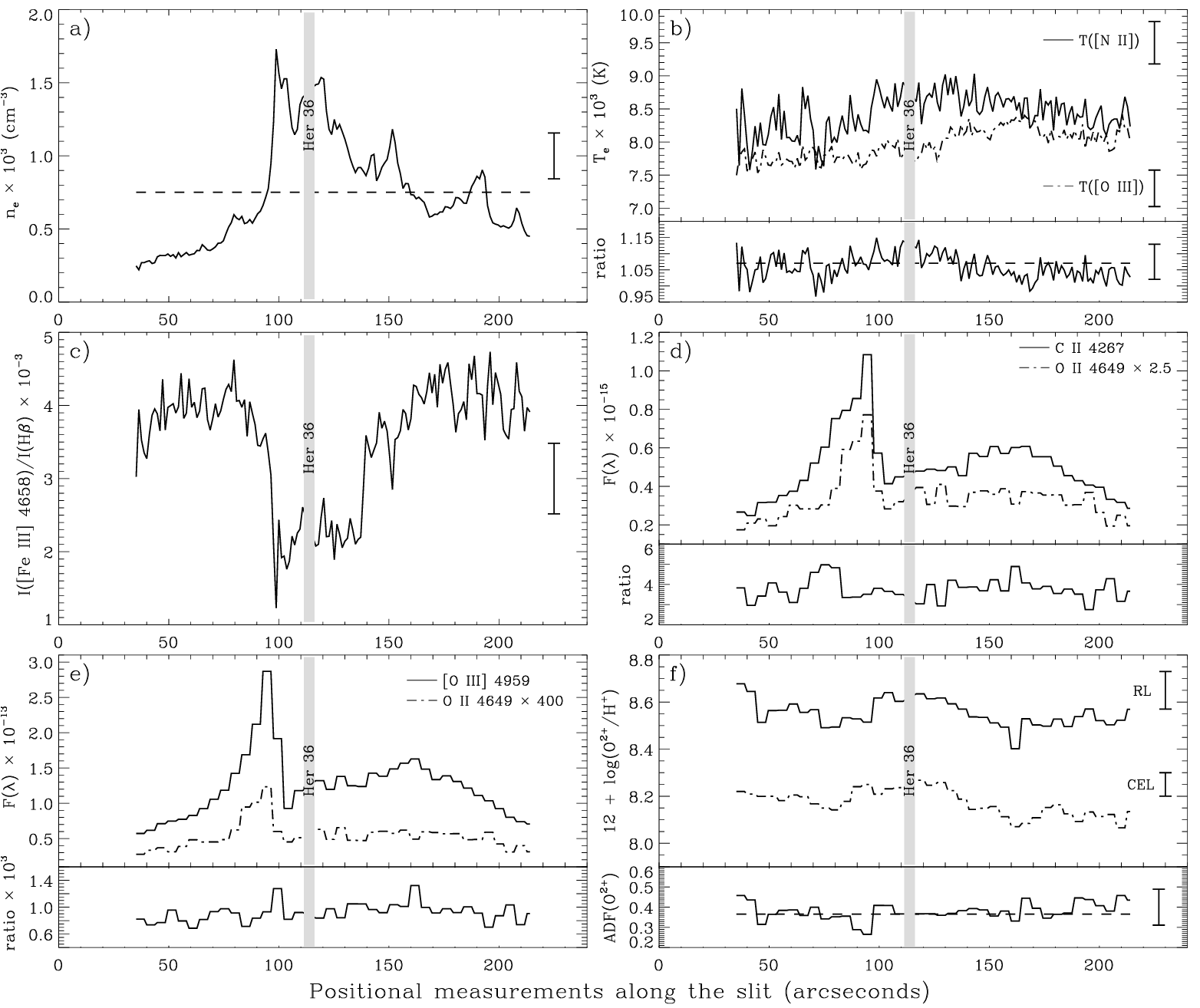} 
   \caption[Spatial profiles of M8 POS2]{Same as Fig.~\ref{m8p1} for the slit position 2 of M8. 
            The grey vertical band covers the apertures affected by stellar features from the 
	    emission of Her~36.}
   \label{m8p2}
  \end{figure*}

  In Figs~\ref{m8p1} and \ref{m8p2} we show the spatial variations of the nebular properties for 
  the slit positions 1 and 2 of M8, respectively. 
  
  The spatial profiles of \nel\ along the slits show a wide range of variation with a maximum 
  associated with the HG region in position 1 (see Fig.~\ref{m8p1}a) and the apertures near 
  Her~36 in the case of position 2 (see Fig.~\ref{m8p2}a). The density peak in Fig.~\ref{m8p1}(a) 
  associated with HH~870 reaches a value higher than 800 \cmc. Following the nomenclature proposed 
  by \cite{ariasetal06}, that peak would be more specifically related to the knot B of this 
  object. At the HG region we find densities between 2600 and 3500 \cmc, values which are in 
  agreement with previous determinations by \cite{garciarojasetal07} and \cite{peimbertetal93b}, 
  who derive densities between 1800 and 3700 \cmc. We have noticed a slight localized enhancement 
  of about 300 \cmc\ in the density profile of position 1 around the positional measurement 
  155\farcs2 (indicated as HH? in Figs~\ref{pos} and \ref{m8p1}). This feature coincides in 
  Fig.~\ref{pos} with a relatively bright rim crossed by the slit and located about 45\arcsec\ to 
  the north of the centre of the HG region. In \S\ref{newhh} we argue that this feature may be a 
  candidate to HH object. In Fig.~\ref{m8p2}(a), the highest density value is related to an 
  ionization front located at 15\farcs3 S and 1\farcs5 E from Her~36. In this slit position, we 
  have also detected local maxima at the positional measurements 65\farcs2, 79\farcs6, 151\farcs6 
  and 192\farcs4. On the one hand, the first two maxima coincide in Fig.~\ref{pos} with gaseous 
  arcs that surround faint stars located approximately at 34\arcsec\ and 49\arcsec\ to the south 
  of Her~36. On the other hand, the other maxima are related to high density filaments located 
  approximately at 40\arcsec\ and 80\arcsec\ to the north of Her~36. The average value of both 
  slit position 1 and 2 obtained from the ``whole slit" spectrum, and included in panel (a) of 
  Figs~\ref{m8p1}-\ref{m8p2} with dashed line, amounts to 1140$\pm$220 and 750$\pm$150 \cmc, 
  respectively.    
 
  The spatial profiles of \te([\ion{O}{iii}]) and \te([\ion{N}{ii}]) are presented in the panels 
  (b) of Figs~\ref{m8p1} and \ref{m8p2} for the slit positions 1 and 2, respectively. We do not 
  find relevant features in these profiles, excluding perhaps very slight enhancement of both 
  temperature indicators at the HG region. However, the temperature rise at these positional 
  measurements are of the order of the error bar. We also present the 
  \te([\ion{N}{ii}])/\te([\ion{O}{iii}]) ratio in the bottom panels of Figs~\ref{m8p1}(b) and 
  \ref{m8p2}(b), which is basically constant along the slits showing spatial variations of the 
  order of the observational errors. In general, \te([\ion{N}{ii}]) is slightly higher than 
  \te([\ion{O}{iii}]), a typical result obtained for \hii\ regions as a consequence of the 
  hardening of the radiation field in the low ionization zones \cite[e.g.][]{stasinska80a}.  
  
  One of the main spectral properties of HH objects is their strong emission in [\ion{Fe}{iii}] 
  lines due to the destruction of dust grains \citep{mesadelgadoetal09b} or changes in the 
  ionization conditions \citep{blagraveetal06}. In Figs~\ref{m8p1}(c) and \ref{m8p2}(c) we have 
  plotted the spatial profile of the dereddened flux of [\ion{Fe}{iii}] 4658 \AA\ with respect to 
  H$\beta$ for the slit positions 1 and 2, respectively. In the slit position 1 we can see a 
  clear rise of the flux at the position of HH~870, which increases by a factor of $\sim$2 with 
  respect to the adjacent background gas. This factor seems to extend along the HG region where 
  the ionization degree decreases and the electron density increases. As in the density spatial 
  profile of slit position 1, we have found in Fig.~\ref{m8p1}(c) an enhancement of the 
  [\ion{Fe}{iii}] emission line at the positional measurement 155\farcs2 where a possible 
  candidate to HH object is located (see \S\ref{newhh}). In the case of the slit position 2, we 
  have found no evidence of important localized increase in the [\ion{Fe}{iii}] emission, only a 
  decrease of a factor of 2 around Her~36.
  
  In Figs~\ref{m8p1}(d)--\ref{m8p2}(d) and \ref{m8p1}(e)--\ref{m8p2}(e) the spatial profiles of 
  the observed flux of \ion{C}{ii} 4267 \AA, \ion{O}{ii} 4649 \AA, and [\ion{O}{iii}] 4959 \AA\ 
  lines along the slit positions are shown, as well as their ratio in the bottom panels of the 
  same figures. It should be reminded that the extractions of the \ion{O}{ii} line are 4\farcs8 
  wide. Then, each extracted aperture represents the average value of four apertures 
  extracted with a spatial size of 1\farcs2. In order to perform a proper comparison, we have 
  also plotted the observed fluxes of \ion{C}{ii} and [\ion{O}{iii}] lines with extractions 
  4\farcs8 wide. The spatial distributions of the pure RLs are quite similar in each slit 
  position. However, we can note a slight rise of the F(\ion{C}{ii} 4267 \AA)/F(\ion{O}{ii} 
  4649 \AA) ratio at the HG and HH~870 regions (see Fig.~\ref{m8p1}d). We have observed a similar 
  enhancement of the ratio of emission lines of species with similar ionization potential to 
  \ioni{C}{2+} (24.4 eV), like \ioni{Ar}{2+} (27.6 eV) or \ioni{Cl}{2+} (23.8 eV), with respect to 
  [\ion{O}{iii}] 4959 \AA\ and \ion{O}{ii} 4649 \AA. In order to explore the reason of this 
  enhancement, we have determined the spatial distribution of the ionization degree which presents 
  an inverse behaviour to the \ion{C}{ii}/\ion{O}{ii} ratio. Though we have not subtracted the 
  telluric emissions from [\ion{O}{ii}] 7320, 7330 \AA\ lines in M8, we have observed that the 
  \ioni{O}{2+}/\ioni{O}{+} decreases between the positional meassurements at 70\arcsec\ and 
  130\arcsec, where the HG region is located, reaching values of about $-$0.4 dex lower than in 
  the rest of the slit position. Therefore, this indicates that the variation in the 
  \ion{C}{ii}/\ion{O}{ii} ratio can be simply due to the decrease of the ionization degree in 
  that zone. On the other hand, the point-to-point comparison of the spatial profiles of 
  [\ion{O}{iii}] 4959 \AA\ and \ion{O}{ii} 4649 \AA\ (see Figs~\ref{m8p1}e--\ref{m8p2}e) does not 
  show clear tendencies.
   
  The [\ion{O}{iii}] spatial distribution of the slit position 1 and 2 of M8 suggests that the 
  gas covered by both slits might not be ionized by the same source. Slit position 1 shows a 
  [\ion{O}{iii}] emission distribution rather similar to the density distribution, but position 2 
  does not. In fact, the [\ion{O}{iii}] emission shows a peak at the ionization front at the south 
  of Her~36 (positional measurement $\sim$95\arcsec) and remains almost constant toward the 
  north of the slit. This area coincides with the dark zone around Her~36. It seems that while the 
  HG region and the area at the south of the positional measurement $\sim$95\arcsec of the slit 
  position 2 are ionized by Her~36, the rest of the position 2 is ionized by another star, perhaps 
  9~Sgr. This hypothesis is consistent with the results from previous works of 
  \cite{woodwardetal86} and \cite{sanchezpeimbert91}.      
   \begin{figure*}
    \centering
    \includegraphics[scale=0.6,angle=90]{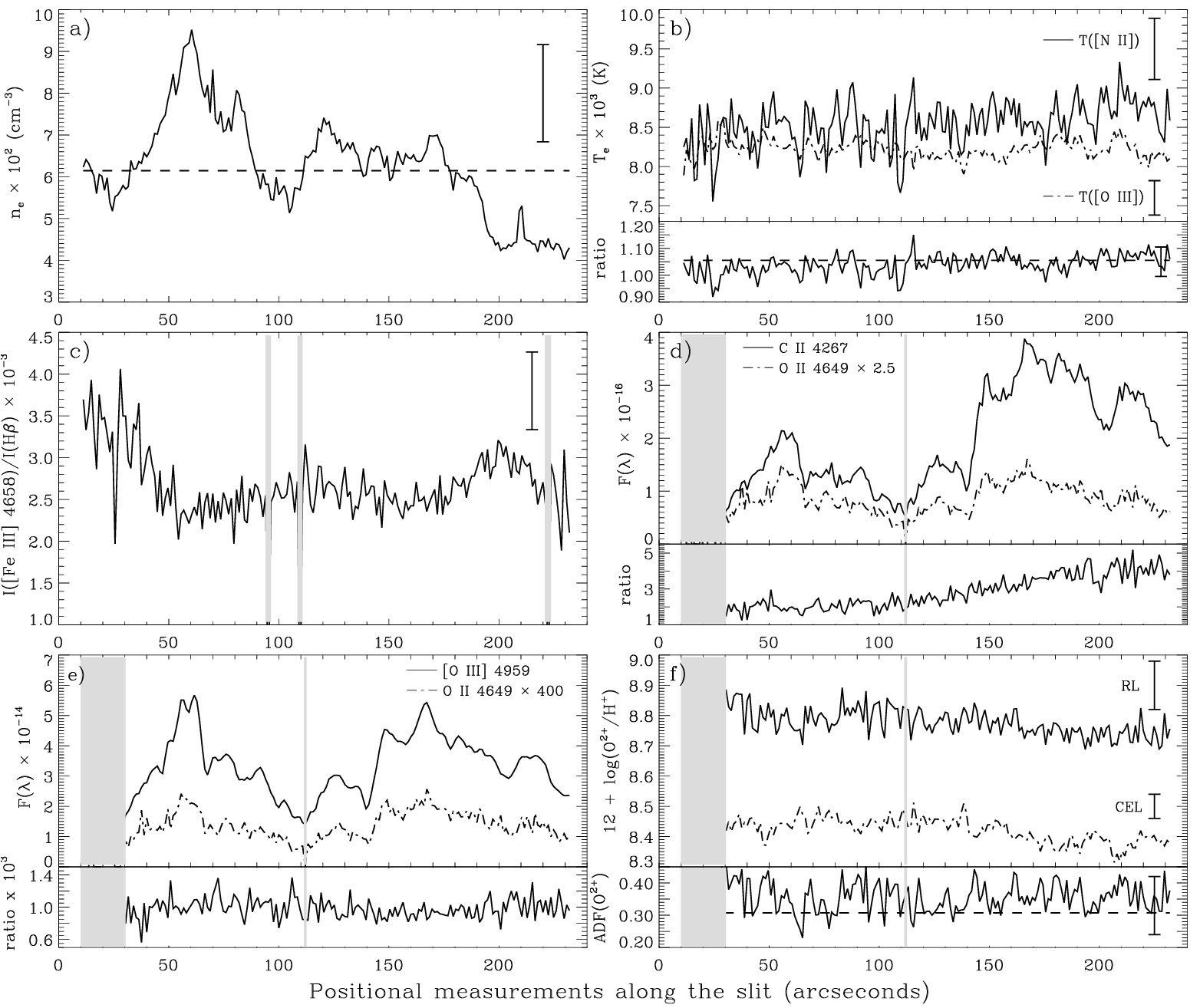} 
    \caption[Spatial profiles of M17]{Same as Fig.~\ref{m8p1} for the slit position of M17. 
             The grey bands cover apertures where [\ion{Fe}{iii}] 4658 \AA\ (c) and \ion{O}{ii} 
    	     4649 \AA\ (d--f) were not detected. Positional measurements along the slit go from 
	     west to east (see Fig~\ref{pos}).}
    \label{m17p1}
   \end{figure*}  
  
  Finally, Figs~\ref{m8p1}(f)--\ref{m8p2}(f) show the spatial variation of the \ioni{O}{2+} 
  abundances derived from RLs and CELs, as well as the AD factor of this ion --\adfo-- defined in 
  its logarithmic form as
  \begin{equation}
    {\rm ADF(O^{2+})} = log({\rm O^{2+}/H^+})_{\rm RLs} - log({\rm O^{2+}/H^+})_{\rm CELs}.
  \end{equation} 
  As in Figs~\ref{m8p1}(d)--(e) and \ref{m8p2}(d)--(e), we present the spatial profile for the 
  extractions 4\farcs8 wide of the \ioni{O}{2+}/\ioni{H}{+} ratio obtained from RLs and CELs. 
  The \adfo\ remains fairly constant along the slit positions 1 and 2 with average values of 
  0.40$\pm$0.11 and 0.37$\pm$0.09 dex, respectively, which are in agreement with previous 
  determinations obtained by \cite{estebanetal99a} --0.34 dex-- and \cite{garciarojasetal07} 
  --0.37 dex. The only relevant variation detected in the \ioni{O}{2+} abundance profiles is at 
  the HG region, where the \ioni{O}{2+} abundance decreases up to 7.6 dex due to the decrease of 
  the ionization degree.  
 \subsection{M17} \label{m17} 
  Fig.~\ref{m17p1} shows the spatial distributions of several nebular properties along the slit 
  position that covers two bright areas of M17 (see Fig.~\ref{pos}). The density profile 
  (Fig.~\ref{m17p1}a) shows a peak between the positional measurements 30\arcsec\ and 90\arcsec, 
  a rather constant value between 90\arcsec\ and 180\arcsec\ and a decrease at the eastern edge. 
  We have found a good agreement between our density determinations and those at the positions 3 
  and 13 studied by \cite{peimbertetal92}, which coincide with our positional measurements 
  60\arcsec\ and 125\arcsec\ (see \S\ref{lir}). In the case of the temperature distributions (see  
  Fig.~\ref{m17p1}b), we only have found variations of the order of our error bar. In the case of 
  the \te([\ion{O}{iii}]), we have found values similar to those derived by \cite{peimbertetal92} 
  in their slit positions. On the other hand, the \te([\ion{N}{ii}]) obtained by those authors 
  shows a higher value at their position 3, which amounts to 11600 K. However, 
  \cite{estebanetal99a} also explored the physical conditions at position 3 of 
  \cite{peimbertetal92} and found a \te([\ion{N}{ii}]) of about 8990 K, which is in agreement 
  with our determinations within the errors. In Fig.~\ref{m17p1}(c), we can find the spatial 
  profile of the [\ion{Fe}{iii}] 4658 \AA\ line flux with respect to H$\beta$, where we do not 
  find any localized enhancement, which may be related to the presence of HH objects.  
   \begin{figure*}
    \centering
    \includegraphics[scale=0.6,angle=90]{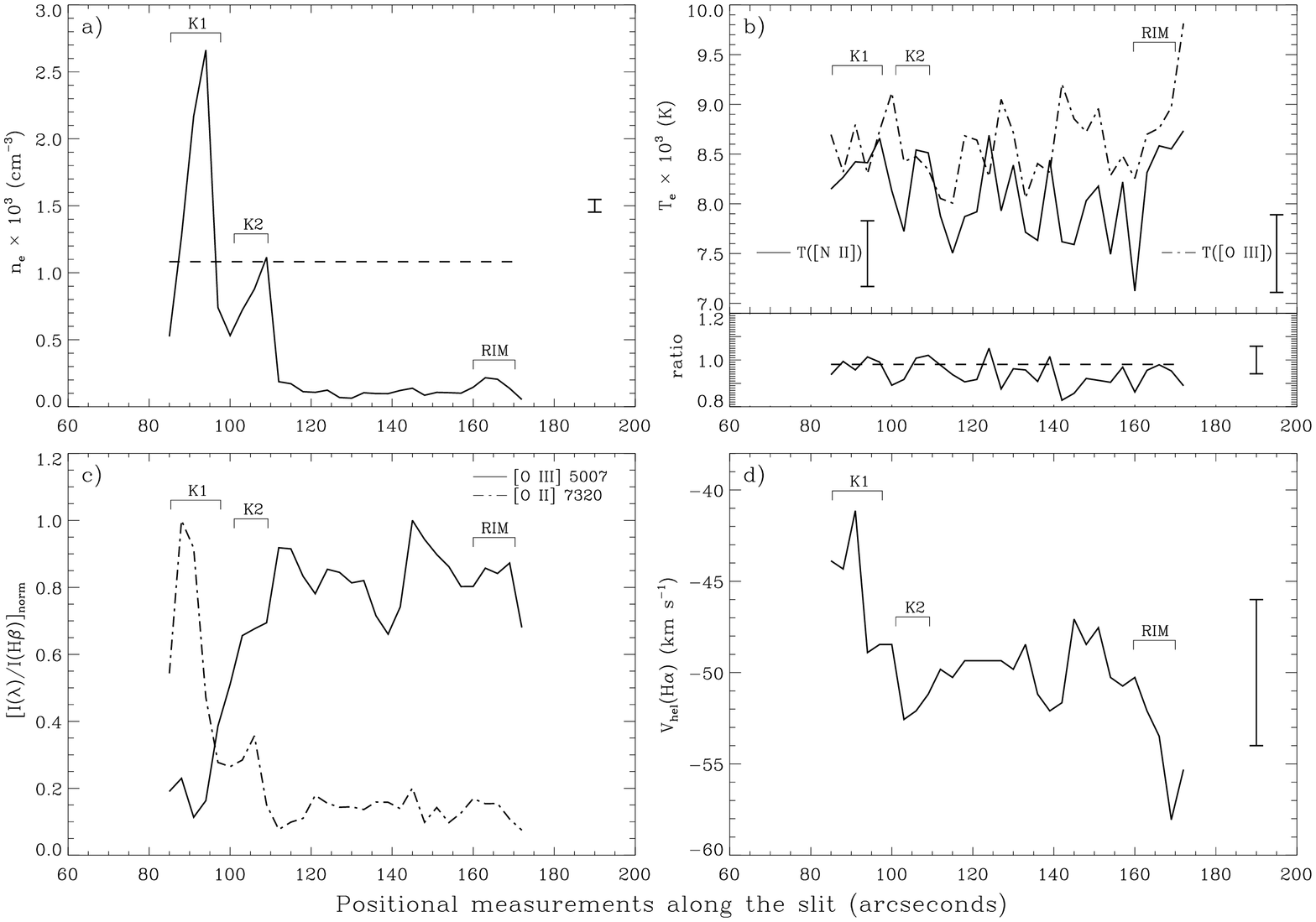} 
    \caption[Spatial profiles of NGC]{Spatial profiles of several nebular properties along 
             the slit position of \ngc. Positional measurements along the slit go from west to 
 	     east (see Fig~\ref{pos}). The position of the knots, K1 and K2, and the RIM are 
 	     indicated as well as the typical error of some variables. The horizontal 
	     dashed line in some panels gives the value obtained from the ``whole slit" spectrum. 
	     (a) Profile of \nel; (b) top: profiles of \te([\ion{N}{ii}]) (solid line) and 
	     \te([\ion{O}{iii}]) (dash-dotted line), botton: profile of 
	     \te([\ion{N}{ii}])/\te([\ion{O}{iii}]) ratio; (c) profile of the dereddened fluxes 
	     with respect to H$\beta$ of [\ion{O}{iii}] 5007 \AA\ (solid line) and [\ion{O}{ii}] 
	     7320 \AA\ (dash-dotted line) normalized to their respective maximum; (d) heliocentric 
	     velocity profile obtained from the centroid of H$\alpha$.}
    \label{ngcp1}
   \end{figure*}  
  
  The spatial distributions of the \ion{C}{ii} 4267 \AA\ and \ion{O}{ii} 4649 \AA\ are presented 
  in Fig.~\ref{m17p1}(d), where we can notice an increase of the \ion{C}{ii} emission flux to the 
  east of the positional measurement 140\arcsec. This rise can be related to a decrease of the 
  ionization degree. In fact, \cite{peimbertetal92} find a change in the \ioni{O}{2+}/\ioni{O}{+} 
  ratio between their slit positions 3 and 13 of about 0.40 dex. The comparison of the 
  [\ion{O}{iii}] and \ion{O}{ii} emission line fluxes in Fig.~\ref{m17p1}(e) does not show 
  relevant features, and their ratio remains essentially constant along the slit.
  
  Finally, Fig.~\ref{m17p1}(f) shows the spatial profiles of the \ioni{O}{2+} abundances derived 
  from RLs and CELs, as well as their ratio, \adfo. As we can see, both abundance determinations 
  show a similar decrease along the slit, while the \adfo\ distribution present a constant value 
  of about 0.37$\pm$0.09 dex. This value is somewhat higher than that obtained in the ``whole 
  slit" spectrum, which amounts to 0.31$\pm$0.07 dex, although the difference is consistent 
  within the errors.   
 \subsection{NGC~7635} \label{ngc}
  The spatial distributions of nebular properties along the slit position of \ngc\ are presented 
  in Fig.~\ref{ngcp1}. Each positional measurement corresponds to an area of 
  3\arcsec$\times$0\farcs98. The results obtained for the integrated spectra  covering 
  separately the bright knots at the west of the central star, K1 and K2, and the rim of the 
  bubble (see Fig.~\ref{pos}) are presented in \S\ref{rlngc}.
  
  In Fig.~\ref{ngcp1}(a) the density profile shows its highest value at the position of knot 
  K1, where it reaches a value of about 2600 \cmc. \cite{rodriguez99b} obtained the density in 
  three zones of \ngc\ and her position 1 coincides with our K1, as well as the slit position 3 
  of \cite{talentdufour79}. \cite{rodriguez99b} derived a density value of about 2800 \cmc, which 
  is in agreement with our results, while  \cite{talentdufour79} derived a density value of about 
  4200 \cmc. Moreover, \cite{mooreetal02b} also obtained a higher density values on the knots and 
  the rim, but with larger uncertainties. Two local maxima can also be seen at the positional 
  measurements corresponding to the knot K2 and the rim. Between the knots and the rim, the 
  density remains rather constant with an average value of about 100 \cmc\ and a standard 
  deviation of about 20 \cmc.
  
  In Fig.~\ref{ngcp1}(b), we can see that the electron temperatures do not show strong variations 
  along the slit and are of the order of the error bar. The average values of both temperatures 
  amount to 8600 K for \te([\ion{O}{iii}]) and 8100 K in the case of \te([\ion{N}{ii}]) with 
  standard deviations of about 390 K and 420 K, respectively. Their ratio shows an average value 
  of 0.94 with a standard deviation of about 0.05. Our temperature determinations are in 
  agreement with the results of \cite{rodriguez99b}, \cite{talentdufour79} and 
  \cite{mooreetal02b}.           

  Fig.~\ref{ngcp1}(c) shows the spatial profiles of the dereddened fluxes of [\ion{O}{iii}] 5007 
  \AA\ and [\ion{O}{ii}] 7320 \AA\ with respect to H$\beta$ and normalized to their respective 
  maximum emission. On the one hand, the [\ion{O}{ii}] spatial profile shows its maximun values at 
  the position of the knot K1 decreasing  progressively up to the positional measurement 
  110\arcsec. On the other hand, the [\ion{O}{iii}] distribution presents an inverse pattern with 
  a progressive increase until the same positional measurement. Beyond this point, both spatial 
  profiles do not show strong variations. A similar result can be seen in the $HST$ images of 
  \cite{mooreetal02b}. The behaviour shown in Fig.~\ref{ngcp1}(c) is due to a dramatic change of 
  the ionization degree. The knots have a lower ionization degree than the gas beyond the 
  positional measurement 110\arcsec, which corresponds to the bubble. 
  
  Finally, Fig.~\ref{ngcp1}(d) shows the heliocentric radial velocity, V$_{hel}$, spatial 
  distribution obtained from the Gaussian fit of the H$\alpha$ line profile. We have obtained the 
  most negative velocity at the position of the rim, which is about $-$54$\pm$4 and in agreement 
  with the mean heliocentric velocity found by \cite{christopoulouetal95} for \ngc. This value is 
  also similar to that of the molecular cloud associated to the bubble and the S162 complex 
  \citep[see][and references therein]{christopoulouetal95}. The expansion of the bubble is very 
  slow, with velocities between 4 and 25 \kms and, therefore, our spectral resolution did not 
  allow us to resolve the line splitting of the expanding bubble. In fact, the velocity profile 
  along the slit shown in Fig.~\ref{ngcp1}(d) does not show any remarkable feature or velocity 
  differences substantially larger than the uncertainties. 
\section{Chemical abundances of selected zones of \ngc} \label{rlngc}
 In this section, we present the physical conditions and ionic and elemental abundances --derived 
 from CELs and RLs-- obtained from the integrated spectra that cover the knots --K1 and K2, the 
 rim of the bubble as well as the ``whole slit" spectrum. The results are shown in 
 Tables~\ref{abngc} and \ref{totngc}.  
  \begin{table*}
   \centering
   \begin{minipage}{120mm}
   \caption{Physical conditions and ionic abundances$^a$ of selected zones of \ngc.}
   \label{abngc}
    \begin{tabular}{cccccc}
     \hline
          &    &    &     & \multicolumn{2}{c}{``Whole Slit"} \\
          & K1 & K2 & RIM &  \tf$=0$&\tf$>0$ \\
     \hline
         \nel([\ion{S}{ii}]) (\cmc)  & 1900$\pm$150 &  890$\pm$70  &  180$\pm$30  & \multicolumn{2}{c}{1080$\pm$90}\\
   \te([\ion{N}{ii}]) (K)     & 8450$\pm$320 & 8150$\pm$300 & 8680$\pm$430 & \multicolumn{2}{c}{8420$\pm$330}\\
  \te([\ion{O}{iii}]) (K)     & 8830$\pm$800 & 8140$\pm$290 & 8640$\pm$320 & \multicolumn{2}{c}{8580$\pm$320}\\
   \hline
         \ioni{C}{2+}$^b$    & 8.21$\pm$0.14 & 8.46$\pm$0.08 & 8.40$\pm$0.12 & \multicolumn{2}{c}{8.44$\pm$0.12}\\
          \ioni{N}{+}        & 7.55$\pm$0.04 & 7.27$\pm$0.04 & 6.87$\pm$0.05 & 7.32$\pm$0.04 & 7.66$\pm$0.06\\
          \ioni{O}{+}        & 8.39$\pm$0.10 & 8.28$\pm$0.10 & 7.97$\pm$0.12 & 8.24$\pm$0.10 & 8.45$\pm$0.17\\
         \ioni{O}{2+}        & 7.32$\pm$0.10 & 8.07$\pm$0.05 & 8.08$\pm$0.05 & 7.89$\pm$0.05 & 8.48$\pm$0.08\\
         \ioni{O}{2+}$^b$    &      --       &       --      &       --      & \multicolumn{2}{c}{8.48$\pm$0.09}\\
          \ioni{S}{+}        & 6.37$\pm$0.04 & 6.06$\pm$0.04 & 5.53$\pm$0.04 & 6.10$\pm$0.04 & 6.43$\pm$0.06\\
         \ioni{S}{2+}        & 6.69$\pm$0.20 & 6.90$\pm$0.09 & 6.71$\pm$0.09 & 6.77$\pm$0.09 & 7.40$\pm$0.12\\
   \hline
\ioni{N}{+}/\ioni{O}{+}      &$-$0.84$\pm$0.11 &$-$1.02$\pm$0.11 &$-$1.10$\pm$0.13 &$-$0.91$\pm$0.11 & $-$0.79$\pm$0.18\\
\ioni{O}{2+}/\ioni{O}{+}     &$-$1.07$\pm$0.15 &$-$0.21$\pm$0.11 &   0.11$\pm$0.13 &$-$0.34$\pm$0.11 & 0.03$\pm$0.19\\
\ioni{S}{+}/\ioni{O}{+}      &$-$2.02$\pm$0.11 &$-$2.23$\pm$0.11 &$-$2.44$\pm$0.13 &$-$2.14$\pm$0.11 & $-$2.02$\pm$0.18\\		

     \hline
    \end{tabular} 
    \begin{description}
      \item[$^a$] In units of 12$+log$(\ioni{X}{+i}/\ioni{H}{+}).  
      \item[$^b$] Determined from RLs.		  
    \end{description}
   \end{minipage}   
  \end{table*}
 
 The physical conditions derived from the ratios of CELs are shown in Table~\ref{abngc}. These 
 values are in agreement with those obtained from the 3\arcsec\ extractions presented in 
 Fig.~\ref{ngcp1}, as well as with the different determinations of the literature that we have 
 cited in \S\ref{ngc}. We can notice that the temperatures calculated from the different 
 indicators are very similar within the errors, in agreement with the result obtained from the 
 temperature spatial distribution along the slit position (see Fig.~\ref{ngcp1}b). The electron 
 density and temperatures found for the ``whole slit" spectra are basically averages of the 
 knots and the rim values.
  \begin{table*}
   \centering
   \begin{minipage}{115mm}
   \caption{Set of ionization correction factors adopted in \ngc.}
   \label{icfs}
    \begin{tabular}{ccccccc}
     \hline
                &            &    &    &     & \multicolumn{2}{c}{``Whole Slit"} \\
       Elements & Unseen ion & K1 & K2 & RIM &  \tf$=0$&\tf$>0$ \\
     \hline 
             C  &   \ioni{C}{+} & 6.20$\pm$1.94 & 2.03$\pm$0.55 & 1.56$\pm$0.50 & 2.34$\pm$0.66 & 1.64$\pm$0.74\\      
      N  &  \ioni{N}{2+} & 1.09$\pm$0.33 & 1.61$\pm$0.43 & 2.28$\pm$0.73 & 1.45$\pm$0.40 & 2.08$\pm$0.93\\ 
      S  &  \ioni{S}{3+} & 1.00$\pm$0.01 & 1.02$\pm$0.01 & 1.07$\pm$0.04 & 1.01$\pm$0.01 & 1.05$\pm$0.05\\

     \hline 
    \end{tabular} 
   \end{minipage}
  \end{table*} 
 
\begin{figure}
    \centering
    \includegraphics[scale=0.65]{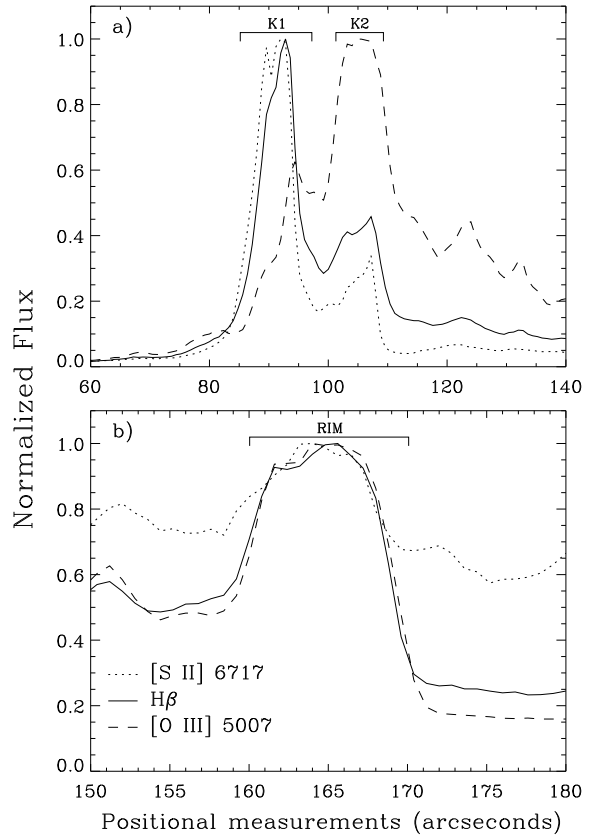} 
    \caption[ngc profiles]{Spatial profiles of several emission lines obtained from extractions 
                           with a size of 0\farcs8 --the average seeing during the observations-- 
			   around the knots (a) and the rim (b).}
    \label{ngcpr}
   \end{figure}

 We have derived ionic abundances of several ions (see Table~\ref{abngc}). The 
 \ioni{O}{+}/\ioni{H}{+} ratio was obtained removing the telluric contamination from the 
 [\ion{O}{ii}] 7320, 7330 \AA\ lines as it was explained in \S\ref{ab}. A proper determination of 
 the \ioni{O}{+} abundances has allowed us to explore the strong variation of the ionization 
 degree between the knots and the rim. As we can see in Table~\ref{abngc}, the ions with low 
 ionization potential, such as \ioni{N}{+}, \ioni{O}{+} and \ioni{S}{+}, show a gradual decrease 
 in their abundance following the sequence: K1, K2 and the rim. As it is expected, this decrease 
 is accompanied by an increase of the abundances of ions of high ionization potential, however 
 the differences between K2 and the rim are not large in this case. In Fig.~\ref{ngcpr} we 
 present the spatial distribution of several emission lines of ions with different ionization 
 potentials: \ioni{S}{+} (10.4 eV), \ioni{H}{+} (13.6 eV) and \ioni{O}{2+} (35.1 eV). The profiles 
 are centered around the knots (Fig.~\ref{ngcpr}a) and the rim (Fig.~\ref{ngcpr}b). As it is shown 
 in Fig.~\ref{ngcpr}(a), the [\ion{O}{iii}] profile shows a completely different behaviour with 
 respect to the other ions showing a progressive decrease from the positional measurement 
 110\arcsec\ towards the west (left in Fig.~\ref{ngcpr}a). This behaviour is produced because 
 most ionizing photons which are able to convert \ioni{O}{+} to \ioni{O}{2+}, are exhausted 
 between the positional measurement 105\arcsec\ and 110\arcsec\ and, therefore, they do not get 
 to penetrate the inner surface of K1, which is facing the ionizing star. Then, K1 is a wall of 
 dense material that maintains the nebula ionization-bounded at that precise location. In the case 
 of K2, the spatial profiles of the three emission lines are rather similar, indicating that this 
 feature is matter-bounded or conversely, it is a ionization front which surface is perpendicular 
 to the line of sight. Finally, the behaviour of the line profiles at the rim --whose surface is 
 tangential to the line of sight-- indicates that it is a matter-bounded feature. This is also 
 consistent with the fact that \te([\ion{O}{iii}]) and \te([\ion{N}{ii}]) are so similar at the 
 rim.
 
\begin{figure}
    \centering
    \includegraphics[scale=0.5,angle=90]{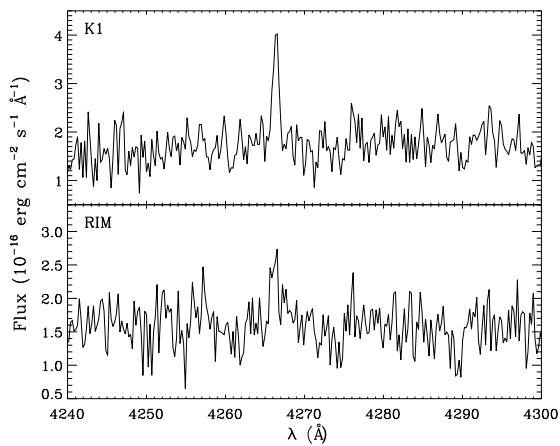} 
    \caption[CII line]{Sections of the integrated spectra around the emission line 
	     \ion{C}{ii} 4267 \AA\ for the knot K1 (top panel) and the rim (botton panel) of 
	     \ngc.}
    \label{ciingc}
   \end{figure}

 It is important to note that we have detected and measured the faint \ion{C}{ii} 4267 \AA\ RL
 (see Fig.~\ref{ciingc}) in all the integrated spectra, as well as five of the eight RLs of the 
 multiplet 1 of \ion{O}{ii} (see Fig.~\ref{rls}d) in the ``whole slit" spectrum, allowing us to 
 derive \ioni{C}{2+} and \ioni{O}{2+} abundances from RLs. Several authors have previously 
 determined the \ioni{C}{2+} abundance in \ngc. \cite{rodriguez99b} obtained values between 8.11 
 and 8.23 dex in her two slit positions which cover the group of knots at the south of K1. On the 
 bubble rim, \cite{mooreetal02b} obtained a \ioni{C}{2+} abundance of 8.63 dex. Our determinations 
 lie in between those values. In addition, \cite{mooreetal02b} obtained the first determination of 
 the \ioni{O}{2+}/\ioni{H}{+} ratio from RLs on the rim bubble, they found a value of 8.47 dex and 
 a moderate \adfo\ $=$ 0.2 dex. However, from our ``whole slit" spectrum we obtain an \adfo\ of 
 about 0.59$\pm$0.10 dex, the highest value found so far in an \hii\ region (see \S\ref{adfb}). 
 
 In order to derive the total gaseous abundances of the different elements, we have corrected for 
 the unseen ionization stages by using a set of ionization correction factors (ICFs). Both adopted 
 ICF values and total abundances are presented in Table~\ref{icfs} and Table~\ref{totngc}, 
 respectively. For carbon, we have adopted the photoionization models of \cite{garnettetal99} to 
 estimate the ICF(\ioni{C}{+}). In the case of oxygen and nitrogen, we considered the classical 
 assumptions O $=$ \ioni{O}{+}~$+$~\ioni{O}{2+} and \ioni{N}{+}/N $=$ \ioni{O}{+}/O, respectively. 
 Finally, we have used the relation proposed by \cite{stasinska78} to derive the ICF(\ioni{S}{3+}) 
 and the S abundances. In general, the total abundances derived for K1 and K2 are rather similar 
 considering the uncertainties --except perhaps in the case of carbon-- but higher than those 
 derived for the spectrum of the rim. For the knots, \cite{rodriguez99b}, \cite{talentdufour79} 
 and \cite{mooreetal02b} determine a total O abundance of about 8.42, 8.72 and 8.78, respectively. 
 Our O abundance determinations for K1 and K2 are only consistent with those of 
 \cite{rodriguez99b}. For the rim, our O/H ratio is somewhat lower than those of 
 \cite{talentdufour79} and \cite{mooreetal02b}, both obtain a 12$+log$(O/H) $=$ 8.49. The largest 
 differences of abundances between the knots and the rim are found for carbon. The relative 
 contribution of \ioni{C}{+} --and therefore the uncertainty of the total C/H ratio-- is far 
 larger in the case of K1. One of the possible reasons of such difference may be the very 
 different ionization degree of the knots and the rim, specially between K1 and the rim, where the 
 \ioni{O}{2+}/\ioni{O}{+} ratio differs in more than 1 dex (see Table~\ref{abngc}). This strong 
 change in the ionization degree, due to the ionization-bounded structure of the knots, involves 
 an important variation in the ICF(\ioni{C}{+}) (see Table~\ref{icfs}). That fact can be affecting 
 the suitability of the this ICF.    
  \begin{table*}
   \centering
   \begin{minipage}{100mm}
   \caption{Elemental abundances$^a$ in \ngc.}
   \label{totngc}
    \begin{tabular}{cccccc}
     \hline
       & & C & N & O & S \\
     \hline
        \multicolumn{2}{c}{K1}           & 9.00$\pm$0.20 & 7.58$\pm$0.14 & 8.42$\pm$0.09 & 6.86$\pm$0.14 \\
        \multicolumn{2}{c}{K2}	         & 8.77$\pm$0.14 & 7.47$\pm$0.12 & 8.49$\pm$0.06 & 6.96$\pm$0.08 \\
        \multicolumn{2}{c}{Rim}	         & 8.59$\pm$0.18 & 7.23$\pm$0.15 & 8.33$\pm$0.06 & 6.77$\pm$0.09 \\	   
\multirow{2}{*}{``Whole Slit"} & \tf$=0$ & 8.80$\pm$0.17 & 7.49$\pm$0.13 & 8.40$\pm$0.08 & 6.87$\pm$0.08 \\			     
                               & \tf$>0$ & 8.65$\pm$0.23 & 7.98$\pm$0.20 & 8.76$\pm$0.09 & 7.46$\pm$0.11 \\
\hline		
\multicolumn{6}{c}{Abundances expected from abundance gradients} \\
\hline	       
\multicolumn{2}{c}{\cite{shaveretal83}}     &     --        & 7.67$\pm$0.15 & 8.60$\pm$0.15 &       --     \\ 
\multicolumn{2}{c}{\cite{afflerbachetal97}} &     --        & 7.76$\pm$0.08 & 8.56$\pm$0.11 & 6.67$\pm$0.07\\  
\multicolumn{2}{c}{\cite{deharvengetal00}}  &     --        &       --      & 8.46$\pm$0.07 &       --     \\	
\multicolumn{2}{c}{\cite{estebanetal05}$^b$}& 8.45$\pm$0.20 &       --      & 8.64$\pm$0.12 &       --     \\	      
\multicolumn{2}{c}{\cite{carigietal05}$^c$} &     --        & 7.74$\pm$0.20 &       --      &       --     \\	      
\hline		
\multicolumn{6}{c}{Solar abundances}\\ 
\hline	
\multicolumn{2}{c}{\cite{asplundetal09}}    & 8.43$\pm$0.05 & 7.83$\pm$0.05 & 8.69$\pm$0.05 & 7.12$\pm$0.03\\	      
     \hline
    \end{tabular} 
    \begin{description}
      \item[$^a$] In units of 12$+log$(X/H).   
      \item[$^b$] Determined from RLs.	
      \item[$^c$] Determined from CELs assuming \tf$>0$. 
    \end{description}
   \end{minipage}
  \end{table*}
\section{Discussion} \label{discu}
 \subsection{The abundance pattern of \ngc} \label{adfb}
  Assuming the validity of the temperature fluctuation hypothesis proposed by \cite{peimbert67} 
  and that this phenomenon is related to the AD problem \citep[see][]{garciarojasesteban07}, we 
  have estimated a \tf\ parameter from the \adfo\ found for the ``whole slit" spectra, following 
  the formalism outlined by \cite{peimbertcostero69} and equations (8)--(11) of 
  \cite{peimbertetal04}. The value of the \tf\ parameter representative of the \ioni{O}{2+} zone 
  is 0.071$\pm$0.009. We have adopted this \tf\ parameter in order to calculate the ionic 
  abundances, the same set of ICFs (see \S\ref{rlngc}) and the total abundances for the 
  ``whole slit" spectrum under the presence of temperature fluctuations. These values are also 
  included in Tables~\ref{abngc}, \ref{icfs} and \ref{totngc}. 
 
  The ``whole slit" abundances can be considered as the representative average values for \ngc\ 
  and, therefore, it seems appropriate to compare them with those expected from the Galactic 
  abundance gradients. We have included in Table~\ref{totngc} the expected abundances obtained 
  from some of the gradients determinations available in the literature, as well as the solar 
  values. Using the distance to \ngc\ of 2.4$\pm$0.2 kpc adopted by \cite{mooreetal02b} and 
  assuming a galactocentric distance of the Sun of 8.0$\pm$0.5 kpc \citep{reid93}, we estimate a 
  galactocentric distance of about 9.2$\pm$0.5 kpc for \ngc. As we can see in Table~\ref{totngc}, 
  the comparison of the total abundances for both possible values of \tf\ and the abundance 
  gradients is rather puzzling. Firstly, the C/H ratio obtained from the \ion{C}{ii} 4267 \AA\ RL 
  is larger than expected by the Galactic abundance gradient of this element derived from 
  observations of \hii\ regions \citep{estebanetal05} for any of the two values of the \tf\ 
  parameter considered in the table. The difference between our C abundance determinations and the 
  expectations of the abundance gradients is specially large in the case or the knots, but 
  marginally consistent within the errors in the case of the rim. This could suggest that --as we 
  outlined in Section~\ref{rlngc}-- this disagreement may be due to the unsuitability of the 
  ICF(\ioni{C}{+}) used, at least in the case of the knots. In the case of the ``whole slit" 
  spectrum, the \ioni{C}{2+}/\ioni{O}{2+} ratio with respect to the excitation 
  degree --\ioni{O}{2+}/\ioni{O}{+} ratio-- of the ionized gas is similar to the values reported 
  for other Galactic \hii\ regions, as S~311 \citep{garciarojasetal05} and, therefore, the 
  ICF(\ioni{C}{+}) seems to be not the reason of the large C/H ratio in this particular case. The 
  expected values of the Galactic O/H and N/H gradients are just in between our determinations for 
  \tf$=$0 and \tf$>0$. In particular, the O abundance expected in the abundance gradient of 
  \cite{estebanetal05} --based on the flux of \ion{O}{ii} RLs-- is marginally consistent with our 
  determinations for \tf$>0$ within the errors. Finally, the S/H ratio calculated for any value of 
  the \tf\ parameter is always larger than that expected from the S gradient of 
  \cite{afflerbachetal97}.  

  The lack of consistency between our abundance determinations and the Galactic abundance 
  gradients expectations for \ngc\ suggests that perhaps the standard methods for deriving 
  physical conditions and chemical abundances are not giving the correct values for this object. 
  The problem seems to affect the spectra of all selected areas: the knots, the rim and the 
  ``whole slit" ones. Firstly, the carbon abundance is too high. It is very unlikely that the 
  nebula has suffered a pollution of C processed by the ionizing central star because it is still 
  in the main sequence. On the other hand, due to the fact that K1 and K2 also show a very intense 
  \ion{C}{ii} 4267 \AA\ RL and high \ioni{C}{2+}/\ioni{H}{+} ratios, the hypothetical chemical 
  pollution event should also affect these structures that belong to the larger emission complex 
  S162 \citep{mooreetal02b}, which is outside the bubble of \ngc. Another possibility of a local 
  increase of C may be due to the destruction of carbon-rich dust by the shock associated with the 
  expanding windblown bubble. A possibility that was already suggested by \cite{mooreetal02b}. 
  \cite{estebanetal98} estimated that the destruction of all the carbon locked-up onto dust grains 
  in the Orion Nebula would increase the measured gas-phase C/H ratio in only about 0.1 dex. An 
  increase of such magnitude would alleviate somehow the difference between the observed values 
  and those expected by the abundance gradients, but does not account for it completely. All these 
  problems with the flux of the \ion{C}{ii} 4267 \AA\ RL and the high C/H ratio, specially at K1, 
  makes us to consider that the emission flux of that line is enhanced by an unknown mechanism. We 
  speculate that perhaps this mechanism could be related with the strong photoevaporative flow 
  from the surfaces of the knots found by \cite{mooreetal02b} from $HST$ images of \ngc. For these 
  authors, structures as K1 are the ionized edge of a mass of neutral material associated to the 
  S162 complex, which is ionized by the central Of-type star. The photoevaporative flow seem to be 
  interacting with the stellar wind as it is suggested by the presence of an emission loop 
  between the knots and the central star \citep{mooreetal02b}. Further deeper and higher spectral 
  resolution spectroscopical observations of these features would be needed to shed light onto 
  this problem. 
 \subsection{Comparison with the Orion Nebula} \label{comp}
  The Orion Nebula is the nearest \hii\ region and therefore the most suitable object to explore 
  the spatial variations of the nebular properties as well as the abundance discrepancy problem at 
  the highest spatial resolution. \cite{mesadelgadoetal08} obtained the spatial distributions of 
  several quantities over five slit positions in the Orion Nebula with an angular resolution set 
  to 1\farcs2. They found significant small-spatial scale variations of density, temperature as 
  well as the \adfo, most of them related to the presence of morphological structures such as HH 
  objects, proplyds and ionization fronts as the Orion bar. 
   
  In this study, we have tried to obtain a similar dataset for other bright --but more distant-- 
  Galactic \hii\ regions. However, the inventories of proplyds and HH objects in M8 
  and M17 are far less complete than for the Orion Nebula. Because of its different nature, there 
  is no detection of these objects in \ngc. Discovering proplyds in other \hii\ regions --apart 
  from the Orion Nebula-- is a difficult task considering their larger distances and the problems 
  related to the discrimation between true proplyds and fragmented portions of molecular clouds 
  \citep[see][]{demarcoetal06}. The single candidates to proplyds in M8 and M17 have been reported 
  by \cite{stecklumetal98} --marked as UC in Fig.~\ref{pos}-- and \cite{demarcoetal06} analyzing 
  narrow-band images taken with $HST$, respectively. On the other hand, HH objects are direct 
  manifestations of the interaction of gas ejected by a young star with its surroundings. The 
  existence of HH objects in M8 was firstly mentioned by \cite{reipurth81}, who discovered HH~213 
  from images with the 1-m telescope class at Las Campanas Observatory. More recently, the catalog 
  of HH objects in M8 has increased with the spectroscopic confirmation of HH~870 
  \citep{ariasetal06} and the new five outflows reported by \cite{barbaarias07} using optical 
  narrow-band imaging with the Wide Field Imager at the 2.2-m telescope at La Silla 
  Observatory. In the case of M17, there is not available literature about the presence of HH 
  objects in this \hii\ region, excluding a possible microjet found by \cite{demarcoetal06}.  

  Assuming a distance to the Orion Nebula of 436 pc \citep{odellhenney08}, the linear sizes of 
  the proplyds and HH objects located at the Huygens region lie between 0.003 and 0.007 pc and 
  0.006 and 0.02 pc, respectively. The extractions analyzed by \cite{mesadelgadoetal08} had an 
  angular size of 1\farcs2, that corresponds to a linear resolution of 0.0025 pc. Given that 
  resolution, \cite{mesadelgadoetal08} could obtain enough spatial sampling to detect the spatial 
  variations of the nebular properties associated to those morphological structures. However, the 
  \hii\ regions we analyse in this paper are at larger distances than the Orion Nebula: 
  1.25 kpc \citep[M8][]{ariasetal06}, 1.6 kpc \citep[M17][]{povichetal07} and 2.4 kpc  
  \citep[\ngc][]{mooreetal02b}. At those distances and with the size of the extractions used for 
  each \hii\ region in this work, the minimum linear size we can resolve is 0.0072 pc, 0.0093 pc 
  and 0.035 pc in the case of M8, M17 and \ngc, respectively. As we can see, the linear 
  resolutions that we achieve in our spatial analysis are a factor 3, 4 and 14, respectively, 
  lower than in the case of the Orion Nebula. It is clear that spatial variations of the physical 
  and chemical properties associated with proplyds with linear sizes similar to those of the Orion 
  Nebula are very unlikely to be detected with our observations, except perhaps in the case of M8, 
  where very large proplyds are of the order of our small extractions. With these comparatively 
  large resolution elements, the emission of the proplyds --and any associated variation of the 
  local properties of the ionized gas, if they exist-- would be diluted with the emission of the 
  ambient nebular gas. In the case of HH objects, structures similar to those observed in the 
  Orion Nebula would be resolved with the size of the extractions used in the cases of M8 and M17. 
  For example, knot B of HH~870 in M8 has a diameter of about 0.015 pc, similar to those found in 
  the Orion Nebula. 

  If those object are present in all \hii\ regions, we have estimated that the spatial resolution 
  needed to resolve proplyds similar to those of the Orion Nebula should be about 0\farcs42 for 
  M8, 0\farcs33 for M17 and 0\farcs22 in the case of \ngc, which are of the order or smaller than 
  the best seeing attainable with ground-based telescopes.       
 \subsection{A possible candidate to Herbig-Haro object in M8} \label{newhh}
  In \S\ref{m8} we have presented several evidences that suggest the presence of a new HH object 
  in M8. It is located approximately 16\arcsec\ east and 44\arcsec\ north from Her~36 and can be 
  seen as a diffuse relatively bright arc near an ionization front (HH? in Fig.~\ref{pos}). This 
  candidate to HH object presents a density 1.4 times larger than the adjacent background, 
  somewhat lower than the density contrast found for the much prominent HH~870, which amounts to 
  2.3.
  
  As we mentioned in \S\ref{m8}, the HH objects typically show a strong emission of lines emitted 
  by low ionization potential ions such as \ioni{Fe}{2+}, \ioni{S}{+} or \ioni{O}{+}. The 
  spatial distribution of [\ion{Fe}{iii}] 4658 \AA\ shown in Fig.~\ref{m8p1}(c) indicates that 
  this line is enhanced by a factor of 1.3 with respect to the surrounding background gas. This 
  increase of the dereddened flux of [\ion{Fe}{iii}] 4658 \AA\ is also lower than that shown by 
  HH~870. The [\ion{S}{ii}]/H$\alpha$ ratio is a good indicator to discriminate between 
  shock-excited and photoionized gas. \cite{ariasetal06} obtained a map of 
  [\ion{S}{ii}]/H$\alpha$ ratio of the central part of M8 nebula, where they reported high values 
  of this indicator at the three nebular knots associated with HH~870. Using our data, we find 
  that the [\ion{S}{ii}]/H$\alpha$ ratio increases by a factor of 1.8 at the position of HH~870. 
  In the case of our candidate to HH object, we have obtained an increment of about 1.3, similar 
  to that we can estimate from the map of \cite{ariasetal06}. In fact the shape of the possible 
  HH object can be well identified in the [\ion{S}{ii}] and [\ion{S}{ii}]/H$\alpha$ maps of 
  \cite{ariasetal06}, while it cannot be noticed in their H$\alpha$ map. This behaviour is  
  different to that we can observe in the ionization front near the candidate, which can be 
  clearly seen in all maps. Further spectroscopic observations at higher spectral resolution would 
  be necessary to study the kinematics of this object in order to confirm its true nature.      
\section{Conclusions} \label{conclu}
 In this article, we have carried out long-slit spectrophotometry at intermediate spectral 
 resolution of the Galactic \hii\ regions M8, M17 and \ngc. The one-dimensional spectra were 
 extracted with a resolution of 1\farcs2 for M8 and M17, and 3\arcsec\ in the case of \ngc. 
 Additional extractions with a spatial size of 4\farcs8 were necessary in order to measure the 
 faint \ion{C}{ii} and \ion{O}{ii} RLs in M8. We have studied the spatial distributions of a large 
 number of nebular quantities along several slit positions covering different morphological 
 structures such as HH objects, ionization fronts or bright knots. The studied quantities were 
 \chb, \nel, \te([\ion{O}{iii}]), \te([\ion{N}{ii}]), the observed and dereddened flux of several 
 emission lines ([\ion{Fe}{iii}] 4658 \AA, \ion{C}{ii} 4267 \AA, \ion{O}{ii} 4649 \AA, 
 [\ion{O}{iii}] 4959,5007 \AA\ and [\ion{O}{ii}] 7320 \AA) and the \ioni{O}{2+} abundances derived 
 from CELs and RLs, as well as the difference of both determinations, \adfo.
 
 The density spatial distributions show a large range of variation across the different slit 
 positions. We have found local maxima associated with the HG region, HH~870, knots and 
 regions with a high surface brightness. On the other hand, the temperature spatial profiles do 
 not show important variations related to the cited structures. The temperatures obtained from 
 the different indicators present the classical behaviour for M8 and M17: those derived from 
 [\ion{N}{ii}] lines are higher than those derived from [\ion{O}{iii}] lines, as expected 
 for ionization-bounded nebulae. In the case of \ngc, both temperatures seem to be very similar 
 considering the error due to the own structure of the Bubble nebula which is matter-bounded. We 
 have also explored the spatial behaviour of the \adfo\ along the slit positions of M8 and M17, 
 which remains rather constant, finding values between 0.3 and 0.5 dex with an average error of 
 about 0.1 dex.
 
 We have analysed the physical conditions and chemical composition of four additional extractions 
 of \ngc: three of them centered on the knots --K1 and K2-- and the rim of the bubble; and the 
 last one corresponding to the ``whole slit" spectrum. On the one hand, the comparison of the 
 \ioni{O}{2+}/\ioni{H}{+} ratio determined from CELs and RLs in the ``whole slit" spectrum 
 produces an \adfo\ of about 0.59 dex. Assuming that the AD problem is related to temperature 
 fluctuations, we have obtained a \tf\ parameter of 0.071. We have found a puzzling pattern in 
 the total abundances derived for \ngc. The total abundances obtained for the knots are slightly 
 higher than those of the rim. The total abundances were compared with those expected by the 
 Galactic abundance gradients, finding that there are discrepancies, specially in the case of C. 
 We suspect that \ion{C}{ii} 4267 \AA\ RL may be abnormally enhanced in \ngc\ due to an unknown 
 physical process, whose investigation is outside the scope of this paper but deserves further 
 more detailed observations.        
 
 Comparing our observations with those of \cite{mesadelgadoetal08}, we conclude that proplyds 
 --and the associated variations of the local properties of the gas-- with linear sizes similar 
 to those found in the Orion Nebula cannot be resolved with the observations reported in this 
 paper. Angular resolutions of the order or smaller than the minimun seeing reached from the 
 ground-based telescope would be needed to distinguish the presence of proplyds.
 
 Finally, we have found several evidences that point out to a possible new candidate to HH object 
 in M8. This new object is located 16\arcsec\ east and 44\arcsec\ north from Her~36 where 
 we have found enhancements in the spatial profile of [\ion{Fe}{iii}] 4658 \AA\ and in 
 [\ion{S}{ii}]/H$\alpha$ map presented by \cite{ariasetal06}. 
\section*{Acknowledgments}
 We are vey grateful to the referee of the paper for his/her comments and suggestions. We thank  
 J. Garc{\'\i}a-Rojas, V. Luridiana and S. Sim\'on-D{\'\i}az for their helpful 
 suggestions. This work has been funded by the Spanish Ministerio de Ciencia y Tecnolog\'\i a 
 (MCyT) under project AYA2004-07466 and Ministerio de Educaci\'on y Ciencia (MEC) under project 
 AYA2007-63030.

\label{lastpage}  

\appendix 
 \section{Emission line fluxes} \label{ap1}
  In Table~A1, we present the aperture number (Column 1), H$\beta$ observed fluxes 
  in units of erg~cm$^2$~s$^{-1}$ (Column 2), dereddened flux line ratios of the main emission 
  lines (Columns 3-14) per slit position in untis of $I$(H$\beta$) $=$ 100 and the average 
  extinction coeficient (last row of each slit position). Several notes should be considered in 
  order to understand Table~A1:
 \begin{itemize}
  \item[$(i)$] The aperture number (Ap) is the identification number of each one-dimensional 
               spectra extracted and it is related to the positional measurement in arcseconds 
               as Ap$_{initial}+$Ap$_{size}\times$Ap. Ap$_{initial}$ corresponds to the 
	       position in arcseconds after the edges of the CCD were discarded for the 
	       extraction of one-dimensional spectra (see extraction procedure in \S\ref{lir}). 
	       Ap$_{initial}$ amounts to 34\arcsec\ in M8 POS1 and POS2, 10\arcsec\ in M17 and 
	       82\arcsec\ in \ngc. Ap$_{size}$ is the extraction size in arcseconds along the 
	       spatial direction (see the selected sizes in \S\ref{lir}).  
  
  \item[$(ii)$] In order to interpret correctly the Ap column in M8 POS2 we should consider that 
                apertures 65, 66, 67 and 68 were removed in this position due to contamination by 
		stellar emission (see \S\ref{lir}).
  
  \item[$(iii)$] We have not included the errors associated with H$\beta$ observed fluxes which 
                 remain between 3\% and 5\% among the different slit positions. 
  
  \item[$(iv)$] We have only listed [\ion{O}{iii}] 4959 \AA\ and [\ion{N}{ii}] 6548 \AA\ 
                nebular lines. [\ion{O}{iii}] 5007 \AA\ and [\ion{N}{ii}] 6583 \AA\ can be 
		obtained from the first ones using their theoretical relation: 
                \begin{equation}
		  I(\text{[O~\sc{iii}] 5007})/I(\text{[O~\sc{iii}] 4959}) = 2.88 
		\end{equation}
		and
                \begin{equation}		   
		  I(\text{[N~\sc{ii}] 6583})/I(\text{[N~\sc{ii}] 6548}) = 2.92.
		\end{equation}	    
		
  \item[$(v)$] In the cases of M8 POS1 and POS2, it should be reminded that additional 
               extractions of 4\farcs8 wide were obtained to achieve proper flux measurements 
	       of \ion{O}{ii} RLs (see \S\ref{lir}). In this sense, an aperture of 4\farcs8 
	       wide represents the average value of four apertures extracted with an angular 
	       size of 1\farcs2. The vertical lines in the left-hand side of the \ion{O}{ii} 
	       RLs cover the four apertures of 1\farcs2 wide. 		
 \end{itemize}  
\end{document}